\documentclass[preprint,showpacs,preprintnumbers,amsmath,amssymb]{revtex4}%
\usepackage{graphicx}
\usepackage{dcolumn}
\usepackage{bm}
\usepackage{amsmath}
\usepackage{amsfonts}
\usepackage{amssymb}%
\providecommand{\U}[1]{\protect\rule{.1in}{.1in}}
\newcommand{\be}{\begin{equation}}
\newcommand{\en}{\end{equation}}
\newcommand{\bea}{\begin{eqnarray}}
\newcommand{\ena}{\end{eqnarray}}
\begin{document}
\title{G-Warm inflation: Intermediate model}
\author{Ram\'on Herrera}
\email{ramon.herrera@pucv.cl}
\author{Nelson Videla}
\email{nelson.videla@pucv.cl} \affiliation{ Instituto de
F\'{\i}sica, Pontificia Universidad Cat\'{o}lica de Valpara\'{\i}so,
Avenida Brasil 2950, Casilla 4059, Valpara\'{\i}so, Chile.}

\author{Marco Olivares}
\email{marco.olivaresr@mail.udp.cl} \affiliation{ Facultad de
Ingenier\'{i}a y Ciencias, Universidad Diego Portales, Avenida Ej\'{e}rcito Libertador 441, Casilla 298-V, Santiago, Chile.}

\date{\today}

\begin{abstract}

A warm-intermediate inflationary universe model is studied in the presence of the Galileon
 coupling $G(\phi,X)=g(\phi)X$. General conditions required for successful inflation
 are deduced  and discussed from the background and cosmological perturbations under slow-roll
  approximation.
 In our analyze we assume that
  the dynamics of our model evolves accordingly two separate regimes,
  namely $3g\dot{\phi}H\gg 1+R$, i.e., when the Galileon term dominates over the standard kinetic term
  and the dissipative ratio, and secondly in the regime where both $3g\dot{\phi}H$ and $R$
  become of the same order than unity. For these regimes and assuming that the coupling parameter
  $g=g_0=$ constant, we consider two
  different dissipative coefficients $\Gamma$; one constant and the other being a function of the inflaton field. Furthermore,
  we find the allowed range in the space of parameters for our  G-warm model by
 considering the latest data of Planck and also the BICEP2/Keck-Array data
 from the  $r=r(n_s)$ plane,
in combination with the conditions in which the Galileon term
dominates and the thermal fluctuations of the inflaton field
predominate over the quantum ones.



\end{abstract}

\pacs{98.80.Cq}
\maketitle




\section{Introduction}

The paradigm of cosmic inflation during the very early universe is
arguably the most successful scenario for explaining several
puzzling features of the Hot Big-Bang theory (HBB), as the
horizon, flatness,  monopole problems, among others
\cite{R1,R102,R106,R103,R104,R105}. One of the most interesting
features of inflation is that it can create primordial
perturbations
\cite{R2,R202,R203,R204,R205}. These primordial perturbations seed
the temperature anisotropies that are observed in the cosmic
microwave background (CMB)
\cite{Aghanim:2018eyx,Ade:2015xua,Ade:2013zuv}, as well as the
observed large-scale structure (LSS) of the universe. Indeed, the
simplest inflation model, which consists in a single field with a
canonical kinetic term and an enough flat potential minimally
coupled to gravity, give predictions that are in agreement with
current observational data
\cite{Akrami:2018odb,Array:2015xqh,Ade:2015lrj,Planck:2013jfk}.

The standard picture of inflation requires two separate phases as follows: first, during the  slow-roll phase,
 the universe undergoes an accelerating expansion during which its energy density is dominated by the potential term of the inflaton
 scalar field. Subsequently, during the reheating phase \cite{Kofman:1997yn,Kofman:1994rk}, the inflaton oscillates around the minimum
 of its potential by dissipating its energy to a radiation bath. Consequently, the universe enters the radiation era of the standard HBB model.
  For comprehensive reviews on several aspects of reheating phase, see Refs.\cite{Amin:2014eta,Allahverdi:2010xz}. An alternative scenario,
  called warm inflation \cite{warm1,warm2}, offers the possibility that the inflaton field dissipates its energy into a radiation bath during the
   slow-phase, triggered by a friction term  added to the background
equations. In this sense, warm inflation is opposed to the
conventional cold inflation avoiding  the reheating stage. In the
framework of warm inflation the Universe smoothly enters the
radiation era, wherewith a reheating phase is no longer required
after the end of inflationary epoch. An useful way to parametrize
the effectiveness of warm inflation is trough the ratio $R \equiv
\Gamma/3H$, where $\Gamma$ denotes the dissipative coefficient
 (or else decay ratio) and $H$ the Hubble rate. The weak dissipative regime
for warm inflation
 corresponds to the condition $R \ll 1$, while $R \gg 1$ characterizes the strong dissipative regime of warm inflation. It is worth to mention that the parameter
 $\Gamma$,
may be computed from first principles in quantum field theory,
taking into account that the microscopic physics resulting from
the interactions between the inflaton and other degrees of freedom
\cite{BasteroGil:2012cm,Bartrum:2013fia,Zhang:2009ge,26,28,PRD}.
In general terms, the decay rate for the inflaton field  may
depend on
 the scalar field itself or the temperature of the thermal bath, or both quantities, or even it can be a constant. Furthermore, thermal fluctuations may play a
  fundamental  role in warm inflation scenario regarding the production of primordial fluctuations \cite{6252602,1126,6252603}. In this sense, the density perturbations arise from
thermal fluctuations of the inflaton  which dominate over the
quantum fluctuations. So that, an essential  condition for warm
inflation to occur is the presence of a radiation component whose
temperature is such that $T>H$, since the thermal and quantum
fluctuations are proportional to $T$ and $H$, respectively
\cite{warm1,warm2,6252602,1126,6252603,6252604,62526,Moss:2008yb,Ramos:2013nsa}.
For a comprehensive review and a representative list of recent
references of warm inflation can be seen in Refs.
\cite{Berera:2008ar,Ramos:2016coz} and
\cite{Das:2018rpg,Motaharfar:2018zyb,Bastero-Gil:2018uep,Li:2018wno,Herrera:2018cgi},
respectively.

In relation to exact solutions for canonical single field
inflation in the  framework of General Relativity (GR), one of the
most appealing comes from a constant potential for the inflaton
field, which yields to de Sitter expansion \cite{R1}. On the other
hand, a power-law dependence of the scale factor in cosmic time,
i.e.
 $a(t)\propto t^p$, where $p>1$, is obtained when an exponential
potential for the inflaton field is introduced \cite{Lucchin:1984yf}. Yet another exact solution
corresponds to intermediate inflation model, for which the scale factor evolves
with cosmic time as follows
\cite{Barrow:1990vx}
\begin{equation}
a(t)=\exp\left[A\,t^f\right], \label{at}
\end{equation}
where $A$ and $f$ are constant parameters, satisfying the
conditions $A>0$ and $0<f<1$. This expansion law becomes slower
than de Sitter inflation, but faster than power-law inflation
instead. Although intermediate inflationary model was introduced
as an exact solution, this expansion gives   a particular scalar
field potential of the type $V(\phi)\propto \phi ^{-4(f^{-1}-1)}$
\cite{Barrow:1993zq}. However,  the predictions of this model,
regarding primordial perturbations, may be studied under the
slow-roll approximation \cite{Barrow:1993zq,Barrow:2006dh}. In
this form, at lowest order in the slow-roll approximation, this
model predicts that the scalar spectral index becomes $n_s=1$ when
$f=2/3$, corresponding to the Harrison-Zel'dovich spectrum, which
is ruled out by current observations. In addition, the predictions
of this model on the $n_s-r$ plane lie outside the joint 95$\%$ CL
contour for any value of $f$
\cite{Barrow:2006dh,Ade:2015lrj,Planck:2013jfk}. It is worth to
mention that,
 the intermediate inflation model can be rescued in the stage of warm inflation thanks to the modified dynamics
 \cite{Kamali:2016frd,Herrera:2016sov,Herrera:2015aja,Herrera:2014nta,Herrera:2014mca,Herrera:2013rra}.

Going further the standard canonical inflaton scenario, there are other single-field models constructed in the framework of Hordndeski
\cite{Horndeski:1974wa}, or generalized Galileon theories \cite{Nicolis:2008in,Deffayet:2011gz,Kobayashi:2011nu,Charmousis:2011bf}, which
is the most general four-dimensional scalar-tensor theories in curved space-time, free of ghosts and instabilities, with second-order equations
of motion.
 Of particular interest is  potential-driven inflation in the presence of a cubic Galileon coupling given by $X\Box \phi$
  (where $X=-g^{\mu \nu}\partial_{\mu}\phi \partial_{\nu}\phi/2$)\cite{DeFelice:2011hq}. Here, the Galileon term may suppress
   the tensor-to-scalar ratio and eventually turn viable some inflationary potentials already discarded by current data in the
    canonical scenario, see e.g., \cite{Tsujikawa:2014rta,Ohashi:2012wf}. Recently, the efforts have been focused in building the
    so-called generalized G-inflation models \cite{Kobayashi:2011nu}, consisting in a general term $G(\phi,X)\Box \phi$, which is included
    to the action for scalar field in addition to the standard kinetic term (for recent references, see
     \cite{Herrera:2018mvo,Ramirez:2018dxe,Maity:2018ipt,Hirano:2016gmv,Unnikrishnan:2013rka}). It is worth to mention
      that the construction of such a model deserves a careful analysis in order to prevent the appearance of instabilities and
      having successful inflation \cite{Kobayashi:2011nu,Ohashi:2012wf,Kamada:2010qe,Kobayashi:2010cm,Burrage:2010cu}, as well as a
      subsequent stage of reheating \cite{BazrafshanMoghaddam:2016tdk}. For instance, the authors in \cite{Ohashi:2012wf} studied
      chaotic and natural
inflation in a Galileon scenario $G(\phi,X)=f(\phi)X$, for two
expressions of the coupling function $f(\phi)$, $f=c/M^3$ and
$f\propto \phi$ discussed in \cite{DeFelice:2011hq,Kamada:2010qe}.
Interestingly, they found that if the Galileon self-interaction
dominates over the standard kinetic term after inflation, the
oscillatory stage of reheating may not take place unless the mass
scales characterizing the several potentials satisfy stringent
constraints in comparison to the canonical case. Alternatively, if
dissipative effects during inflation are taken into account, is
possible to study the dynamics of warm inflation scenario in the
presence of a Galileon term. This possibility was addressed first
in Ref.\cite{Herrera:2017qux}, and subsequently following the same line for the thermal fluctuations in
Ref.\cite{Motaharfar:2017dxh}.
Particularly, in \cite{Herrera:2017qux}, it was studied the
Galilean term $G(\phi,X)=g(\phi)X$, when the coupling constant $g$
and the decay rate $\Gamma$ are constant. Here,  considering the
exponential potential, it was found the possibility of distinguish
pure warm inflation or pure generalized G-inflation from the background and of
the
thermal fluctuations. In addition, the modified dynamics may yield
a tensor-to-scalar ratio much smaller than those obtained in a
standard G-inflation scenario, see e.g.,
\cite{Herrera:2017qux,Unnikrishnan:2013rka}.

Regarding the viability of the intermediate inflation in G-inflation scenarios for the cold models, in Refs.\cite{Teimoori:2017jzo} and \cite{Herrera:2018ker},
the authors studied
the inflationary dynamics for such an expansion law for a Galileon term $G(\phi,X)\propto X^n$ and
$G(\phi,X)\propto \phi^{\nu}X^n$, respectively.
For both Galileon couplings, it was found the importance of the power $n$ in order to make compatible the intermediate
inflation model  with current observations.
 In particular, the authors in \cite{Herrera:2018ker} found that for $n>38$ the tensor-to-scalar ratio becomes compatible with the bound $r_{0\textup{.}05}<0\textup{.}07$ (95 $\%$ CL),
 set by the BICEP2/Keck-Array collaboration \cite{Array:2015xqh}. So that, intermediate
 inflation in the framework of  cold model is still rule out for the Galileon term $G(\phi,X)\propto X$ ($n=1$) .

In this form, the main goal of the present paper is to explore the
viability of the intermediate model in the context of the  warm
inflation scenario in which  the Galileon term is given by
$g(\phi)X$. In doing so, we consider a constant coupling function
$g_0$ and in order to parametrize the dissipative effects, we
consider two several expressions for the decay rate:
$\Gamma=\Gamma_0$ and $\Gamma (\phi) \propto V(\phi) $,
respectively. Thus, for each expression of the parameter
$\Gamma$, we will be studied the background as well perturbative
dynamics  for two separate regimes. Firstly, we will consider the regime  in which  the quantity
$3g_0\dot{\phi}H\gg 1+R$, i.e., when the Galileon term dominates
over the standard kinetic term and the dissipative ratio.
Secondly we will analyze  the regime where both quantities $3g\dot{\phi}$
and $R$ become of the same order than unity. For all the cases, we
will obtain the allowed range in the space of parameters. In this
sense, we will  consider the condition for warm inflation $T>H$,
the conditions for the regimes
 $3g_0\dot{\phi}H\gg 1+R$ and $R\sim 3g_0\dot{\phi}H\sim 1$, respectively, together
  with the constraints on the $n_s-r$ plane by latest observational  data.

The  paper is organized as follows: The next section presents a general set up of warm inflation scenario in the presence of a
Galileon term $G(\phi,X)=g(\phi)X$ at background level as well as perturbation level, where expressions for the most relevant cosmological observables
 as the power spectrum of scalar perturbations, scalar spectral index, and the tensor-to-scalar ratio will be obtained. Subsequently, in Section \ref{section40},
 the background and perturbative dynamics for our concrete intermediate inflation will be study in the dominated Galileon regime for
 $\Gamma=\Gamma_0$ and $\Gamma (\phi)\propto V(\phi)$, respectively. Section \ref{section41} is devoted to study the dynamics of our model evolving according
 to the general regime $R\sim 3g_0\dot{\phi}H\sim 1$, also for the cases in which
  $\Gamma=\Gamma_0$ and $\Gamma (\phi)\propto V(\phi)$, respectively. Finally, Section \ref{conclu},
 summarizes our results and presents our conclusions. We use units in which $c=\hbar=M_p$=8$\pi$=1.

\section{G-Warm inflation: Basic equations.\label{secti}}
In this section we give a brief review on  the scenario of G-warm
inflation. We start by writing down
 the 4-dimensional action  for this model

\be S=\int\sqrt{-g_4}\left( {R \over 2}+K(\phi, X)-G(\phi,
X)\,\Box \phi\right) d^4x +S_{\gamma}+S_{int} \,.\label{accion}
\en Here the quantity  $g_{4}$  denotes the determinant of the
space-time metric $g_{\mu\nu}$, $R$ corresponds to the Ricci
scalar, $\phi$  denotes the scalar field and
$X=-g^{\mu\nu}\partial_{\mu}\phi\partial_{\nu}\phi/2$. Besides,
the quantities $K$ and $G$ are arbitrary functions of $X$ and the scalar
field $\phi$. Additionally, we consider that the action for
the perfect fluid describing radiation
is defined by $S_\gamma$ and the interaction action is given
 by $S_{int}$. In this context, $S_{int}$
corresponds to the interaction between the scalar field and other
degrees of freedom \cite{Herrera:2017qux,nw1,nw2}.

By assuming a spatially  flat Friedmann-Robertson-Walker (FRW)
metric,  the Friedmann equation can be written as
\begin{equation}
3 H^{2}=\,\rho=\,[\rho_{\phi}+\rho_{\gamma}],
\label{HC}%
\end{equation}
where the total energy density $\rho$ is given by
$\rho=\rho_{\phi}+\rho_{\gamma}$, whit $\rho_{\phi}$ corresponding
to the energy density of the scalar field $\phi$ and
$\rho_{\gamma}$ denotes  the energy density of the radiation
field, respectively.

Following Refs.\cite{DeFelice:2011hq,Kobayashi:2010cm}, we can identify that the energy density and pressure related to the
scalar field from the action (\ref{accion}) are given by \be
\rho_{\phi}=2K_{X}X-K+3G_{X}H\dot{\phi}^3-2G_{\phi}X
\,,\label{frwa} \en and \be
p_{\phi}=K-2(G_{\phi}+G_{X}\ddot{\phi})X \,,\label{frwb} \en
respectively. In the following, we will consider a homogeneous scalar field, i.e. $\phi=\phi(t)$ and  the subscript $K_X$ corresponds to
 $K_X=\partial K/\partial X$, $G_\phi$ to $G_\phi=\partial G/\partial\phi$,  $K_{XX}=\partial ^2K/\partial X^2$, and so on.

As it was already mentioned, in the scenario of  warm inflation, the universe
is filled with a self-interacting scalar field and a radiation fluid.
In this context,
  the dynamical equations for the densities
$\rho_{\phi}$ and $\rho_{\gamma}$  can be written as \cite{warm1,warm2}
\begin{equation}
\dot{\rho_{\phi}}+3\,H\,(\rho_{\phi}+p_{\phi})=-\Gamma\;\;\dot{\phi}^{2},
\label{key_01}%
\end{equation}
and
\begin{equation}
\dot{\rho}_{\gamma}+4H\rho_{\gamma}=\Gamma\dot{\phi}^{2}. \label{key_02}%
\end{equation}
Here,  we emphasize that the coefficient $\Gamma>0$ corresponds to
the dissipation coefficient and its dependence can be considered
to be a function of the temperature of the thermal bath $T$, in
which $\Gamma(T)$, or the scalar field $\Gamma(\phi)$, or both
$\Gamma(T,\phi)$ or simply a  constant\cite{warm1,warm2}. Recall that, the role of the coefficient $\Gamma$ is to account of
the decay of the scalar field into radiation during the
inflationary stage.

From Eqs.(\ref{frwa}) and (\ref{frwb}) we can rewrite
Eq.(\ref{key_01}) as

$$
3\dot{H}G_{X}\,\dot{\phi}^2+\ddot{\phi}\left[3HG_{XX}\,\dot{\phi}^3
-\dot{\phi}^2(G_{\phi
X}-K_{XX})+6HG_{X}\,\dot{\phi}-2G_{\phi}+K_{X}\right]
$$
\be +3HG_{\phi X}\,\dot{\phi}^3+\dot{\phi}^2(9H^2G_{X}-G_{\phi
\phi }+K_{\phi X})-K_{\phi} -3H\dot{\phi}(2G_{\phi
}-K_{X})=-\Gamma\dot{\phi} \,.\label{frwc} \en

In order to study  our model in the G-warm inflation scenario,
we will consider   the specific  case  in which  the functions
$K(\phi,X)$ and $G(\phi,X)$ are given   by

\be K(\phi, X)=X-V(\phi), \qquad \text{and}  \qquad G(\phi,
X)=g(\phi)\,X \,,\label{kyg} \en where, the quantity  $V(\phi)$
denotes  the effective  potential and the coupling parameter $g$ is a
function that only depends on the scalar field i.e.,
$g=g(\phi)$.

In the context of warm inflation, the energy density related to
the inflaton field $\rho_\phi$ dominates over the energy
density of the radiation field $\rho_\gamma$ during the inflationary epoch, wherewith
$\rho_\phi\gg\rho_\gamma$
\cite{warm1,warm2,6252602,1126,6252603,6252604,62526}. Also, considering the
slow roll approximation in which the effective potential $V(\phi)$
dominates over the functions $X$, $|G_{X}H\dot{\phi} ^3|$ and
$|G_{\phi}X|$, see e.g. \cite{Kobayashi:2010cm}, then   the Friedmann
equation given, by Eq.(\ref{HC}), is reduced to
\begin{equation}
3H^{2}\approx\rho_{\phi}\approx V(\phi). \label{inf2}%
\end{equation}

By assuming the  slow-roll approximation, we can also
 introduce  the set of slow-roll parameters for G-inflation,
 defined
 as \cite{Kobayashi:2010cm}

\be \varepsilon_1={(-\dot{H})\over H^2}, \quad
\epsilon_2={(-\ddot{\phi})\over H\dot{\phi}}, \quad
\epsilon_3={g_{\phi}\dot{\phi}\over g H}, \quad \text{and} \quad
\epsilon_4={g_{\phi \phi}X^{2}\over V_{\phi}} \,.\label{srp} \en

In this sense,  after replacing the functions $K$ and $G$ given by
Eq.(\ref{kyg}), together with the set of  slow roll parameters given by
Eq.(\ref{srp}), we rewrite   the equation of  motion for $\phi$
given  by (\ref{frwc})  as follows \be
3H\dot{\phi}(1-\epsilon_2/3+R)+
3gH^2\dot{\phi}^2[3-\varepsilon_1-2\epsilon_2+2\epsilon_2\epsilon_3/3]
=-V_{\phi}(1-2\epsilon_4) \,.\label{frwc2} \en Here, $R$ denotes
the ratio between $\Gamma$ and the Hubble rate and it is defined as
$R=\frac{\Gamma}{3H}$.

Thus, under  the slow-roll approximation in which the  parameters
$|\varepsilon_1|, |\epsilon_2|, |\epsilon_3|, |\epsilon_4|\ll 1$,
 we obtain that the slow-roll equation of motion for the inflaton field (\ref{frwc2}) is
 reduced to \cite{Herrera:2017qux}

\be 3H\dot{\phi}(1+R+\mathcal{A}) \simeq-V_{\phi} \,,\label{campo}
\en where the function $\mathcal{A}$ is defined as
$\mathcal{A}=3\,g(\phi)H\dot{\phi}$. From the Friedmann equation
(\ref{inf2}), we find that  the Eq.(\ref{campo}) can be rewritten
as \be
 \dot{\phi}^2(1+R+\mathcal{A}) \simeq2(-\dot{H}) \,.\label{gfi}
  \en

For the radiation field, we assume   that during the stage of warm
inflation, the radiation production is quasi-stable, implying that
$\dot{\rho
}_{\gamma}\ll4H\rho_{\gamma}$ and $\dot{\rho}_{\gamma}\ll\Gamma\dot{\phi}^{2}%
$\cite{warm1,warm2,6252602,1126,6252603,6252604,62526}. In this form, during inflation, Eq.(\ref{key_02}) becomes
\begin{equation}
 \rho_{\gamma}\simeq
\frac{\Gamma\dot{\phi}^{2}}{4H}.\label{gamma8}
\end{equation}

We note that the energy density $\rho_\gamma$ and the temperature
of the  thermal bath $T$ are related through
 $\rho_{\gamma}=C_{\gamma}\,T^{4}$, where  $C_{\gamma}%
=\pi^{2}\,g_{\ast}/30$ and   $g_{\ast}$ corresponds to  the number of
relativistic degrees of freedom. Thus, the temperature of the
thermal bath, considering  Eq.(\ref{gamma8}) can be expressed as
\begin{equation}
 T\simeq\left[ \frac{\Gamma\dot{\phi}^{2}}{4C_{\gamma}H}\right]
^{1/4}.\label{temp}
\end{equation}

In G-warm inflation, one may distinguish several regimes, see ref.\cite{Herrera:2017qux}.
From the slow-roll  equation given by Eq.(\ref{campo}), the regimes $R + 3gH\dot{\phi} \ll 1$
and $1 + 3gH\dot{\phi} \ll R$ are the standard weak and
strong dissipative regimes in the scenario of warm inflation for a canonical scalar field, respectively. Now, in
G-warm inflation we can also have
  the regime $1+R \ll| gH\dot{\phi} |$,where the Galileon coupling
dominates during the inflationary epoch and therefore  the dynamics of standard or pure warm inflation
is modified. Also, another two interesting regimes were studied in
ref.\cite{Herrera:2017qux}. Here, the standard weak and strong dissipative regimes
are mixed with the Galileon effect, and these correspond to
 $R \ll1 + 3gH\dot{\phi}$ and $1 \ll R + 3gH
\dot{\phi}$, respectively.

At background level,
another important quantity is the number of $e$-folds
$N$ between two different values of cosmological times
$t_{1}$ and $t_{2}$, defined as $N=\int_{t_{1}}^{t_{2}}\,H\,dt$. In particular for intermediate inflation, $N$ is given by
\begin{equation}
N=\int_{t_{1}}^{t_{2}}\,H\,dt=A\,\left(
t_{2}^{f}-t_{1}^{f}\right).
\label{N10}%
\end{equation}
In this sense, we noted that the Hubble rate assuming the
intermediate expansion   can be expressed in terms of the $e$-folds
$N$ as follows
\begin{equation}
H(N)=Af\,\left[{Af\over 1+f(N-1) } \right]^{1-f\over f},\label{HN}
\end{equation}
and $\dot{H}=\dot{H}(N)$ as
\begin{equation}
 \quad
-\dot{H}(N)=Af(1-f)\,\left[{Af\over 1+f(N-1) } \right]^{2-f\over f}. \label{DHN}%
\end{equation}
Here,  we have considered that the inflationary scenario begins at
the earliest possible stage in which
$\varepsilon_1(t=t_1)=-\dot{H}/H^2=1$\cite{Barrow:1990vx,Barrow:1993zq}.  We also mentioned that
during intermediate expansion, the slow-roll parameter
$\varepsilon_1$ in terms of the number of $e$-folds $N$ becomes
\begin{equation}
\varepsilon_1=-\frac{\dot{H}}{H^2}=\frac{1-f}{1+f(N-1)}.\label{Nf}
\end{equation}
This suggests that the inflationary epoch begins at the earliest
possible stage when the number of $e$-folding is equal to $N=0$. or equivalently  $\varepsilon_1\equiv 1$.
Note that when $N\gg 1$, the slow-roll parameter $\varepsilon_1\rightarrow 0$, implying that inflation never ends.
However, in the context of warm inflation the universe smoothly enters to the
radiation era, since the radiation field dominates over the energy density of
the
inflaton  according as  the universe expands \cite{warm1,warm2},
see also Ref.\cite{sh}  as other mechanisms for address the end of the accelerated expansion and  the reheating of the universe or this expansion law.

On the other hand, the cosmological perturbation theory in the model of
G-warm inflation was developed in Ref.\cite{Herrera:2017qux}. In this context, the
source of the density fluctuations corresponds to thermal
fluctuations of the inflaton field during inflation. Thus, according to   the
evolution of warm inflation, the fluctuations of the inflaton
field $\delta\phi$ are dominantly thermal rather than quantum, see
refs. \cite{warm1,warm2,6252602,1126,6252603,6252604,62526,Moss:2008yb,Ramos:2013nsa}. In order to determine the amplitude of the fluctuations is necessary to consider
 the Langevin equation that includes a
thermal stochastic noise term in the KG equation. In  this way,  the fluctuations of the scalar field
$\delta\phi$ in G-warm model for the case in which the dissipation coefficient
$\Gamma=\Gamma(\phi)$, can be written as
$\delta\phi^2\simeq\sqrt{3H^2+H\Gamma+18gH^3\dot{\phi}}\,\,\,T/2\pi^2$,
see ref.\cite{Herrera:2017qux}. Here, we  noted that in the limit $g\rightarrow 0$,
the fluctuations of the scalar field $\delta\phi$ reduces to the
fluctuations found in the case of pure warm inflation \cite{warm1,warm2,6252602,1126,6252603,6252604,62526,Moss:2008yb,Ramos:2013nsa}. In this form,
following \cite{Herrera:2017qux}, the power spectrum of the
scalar perturbation  defined by
${\mathcal{P}_{\mathcal{R}}}=(H/\dot{\phi})^2\delta\phi^2$, can be
written as \be
 {\mathcal{P}_{\mathcal{R}}}={1\over 2\pi^2 }
 \left({H\over \dot{\phi}} \right) ^2
 \left[
 {\Gamma\,X\over 2C_{\gamma}\,H}  \right]
 ^{1/4}
 \sqrt{3H^2(1+6gH \dot{\phi})+\Gamma\,H}.\label{Py}
\en By using the fact that the rate $R=\Gamma/3H$ and the function
$\mathcal{A}=3\,g \,H\dot{\phi}$, then the scalar perturbation
${\mathcal{P}_{\mathcal{R}}}$ can be rewritten as
 \be
 {\mathcal{P}_{\mathcal{R}}}={\sqrt{3}\over 2\pi^2}
 \left(
 {3\over 4C_{\gamma}} \right)
 ^{1\over 4}
 \left( H^3R ^{1\over 4}\right)
 \dot{\phi}^{-{3\over 2}}
 \sqrt{1+R+2\mathcal{A}}
 \,.\label{PRy2}
 \en

As the scalar spectral index $n_s$ is given by $n_s-1=\frac{d\ln
\,{\cal{P}_{R}}}{d\ln k}$, we find that the spectral index $n_s$
results
$$
n_s\simeq1-
\frac{\epsilon_1}{2}\left[\frac{7}{2}+\frac{1+12gH\dot{\phi}}{(1+R+6gH\dot{\phi})}\right]
+3\epsilon_2\left[\frac{1}{4}-\frac{gH\dot{\phi}}{(1+R+6gH\dot{\phi})}\right]
+\epsilon_3\left[\frac{3gH\dot{\phi}}{(1+R+6gH\dot{\phi})}\right]
$$
\begin{equation}
+\frac{\epsilon_5}{2}
\left[\frac{1}{2}+\frac{R}{(1+R+6gH\dot{\phi})}\right], \label{ns}
\end{equation}
where the quantity  $\epsilon_5$ is defined as $
\epsilon_5=\left(\frac{\dot{\phi}}{H}\right)\,\left(\frac{\Gamma_\phi}{\Gamma}\right).
$ Here, we have used Eq.(\ref{PRy2}).

It is well known that tensor perturbations  during inflation
would generate gravitational waves (GWs). In the case of
G-inflation, the amplitude   of the tensor perturbations is the same
as in the case of  standard general relativity (GR)\cite{DeFelice:2011hq,Kobayashi:2010cm}. So that, the the amplitude
of the tensor perturbations is given by
\begin{equation}
{\cal{P_G}}={2H^2\over \pi^2}.\label{onda}
\end{equation}
Here, we have considered the slow-roll approximation given by
Eq.(\ref{inf2}).

Another important cosmological
observable is the tensor-to-scalar ratio $r={\cal P_{G}}/\cal{P_{R}}$. Thus, from Eqs.(\ref{PRy2}) and
(\ref{onda})  the tensor- scalar ratio can be written as
\begin{equation}
r=4\,X\,\left(\frac{2C_\gamma\,H}{\Gamma\,X}\right)^{1/4}\,[3H^2(1+6gH\dot{\phi})+H\Gamma]^{-1/2}\;.\label{te}
\end{equation}

In the following, we will study the intermediate expansion in the
framework of G-warm inflation, for the simplest case in which the
Galileon coupling function $g=g_0=$
constant\cite{DeFelice:2011hq,Kobayashi:2010cm}. Also, in this
framework   we will  consider two different dissipative
coefficients $\Gamma$. As well, we will restrict ourselves to the
domination of the Galileon effect on standard warm inflation,
i.e., $ 3gH\dot{\phi}={\cal{A}}\gg 1+R$ and we will also studied
the regime where all terms of Eq.(\ref{campo}) are the same
order i.e., $1\sim R\sim \cal{A}$, namely the general
or full solution.

\section{ Domination of the Galileon  regime ${\mathcal{A}}\gg1+R$.\label{section40}}

In this section we utilize the formalism of above to  G-warm
inflation in the context of intermediate expansion, assuming that
our G-warm model evolves according to the domination of the  Galileon
regime, in which the function ${\mathcal{A}}\gg 1+R$.

By assuming the limit ${\mathcal{A}}\gg 1+R$, we note that the background
equations do not depend on
 the dissipation coefficient $\Gamma$. In this way,  we find
 that the speed of scalar field
$\dot{\phi}$ given by Eq.(\ref{campo})
 results in \be \dot{\phi}
\simeq\left[ {2(-\dot{H})\over 3\,g_0 H} \right] ^{1/3}.
\label{gfi1} \en
As we mentioned above, we observed  that $\dot{\phi}$ does not depend of the
coefficient $\Gamma$. Now,
from the intermediate scale factor given by
Eq.(\ref{at}), we obtain that the solution for the scalar field in
terms of the cosmological time becomes

\begin{equation}
\phi(t)=\left[ {9(1-f)\over 4\,g_0}  \right] ^{1/3}
t^{2/3}+C_0, \label{wr11}%
\end{equation}
where $C_0$ denotes an integration constant, that
without loss of generality it can be assumed $C_{0}=0$. From this
solution, we find that the Hubble rate
has the following dependence on the inflaton field
\begin{equation} H(\phi) =A\,f\,\left[ \frac{3}{2} \left({1-f\over g_0}\right)^{1\over 2}\right]
^{(1-f)} \phi^{-{3(1-f)\over 2}}.\label{H1}
\end{equation} In this
way, from Eqs.(\ref{inf2}) and (\ref{H1}) we obtain that the
effective potential in limit  ${\mathcal{A}}\gg 1+R$ is given by
\begin{equation}
V(\phi)=V_0\,
\phi^{-3(1-f)},\,\,\,\;\;\;\;\mbox{where}\;\,\,\,\,V_0=3A^2\,f^2\,\left[
\frac{3}{2} \left({1-f\over
g_0}\right)^{1\over 2}\right] ^{2(1-f)}. \label{pot11}%
\end{equation}

 Note that this kind of
scalar potential (power-law), which depends on the inflaton field in an inverse power-law way, does not have a minimum and it decays to zero for
lager values of $\phi$, since $0<f<1$. We also note that this potential becomes
independent of the dissipation coefficient $\Gamma$, as it was previously quoted.



On the other hand, the dimensionless slow-roll parameter
 $\varepsilon_1=-\dot{H}/H^2$ can be rewritten in terms of the inflaton field,
considering the slow-roll approximation wherewith
$$
\varepsilon_1=\left( {\frac{1-f}{Af}}\right) \left[ \frac{3}{2}
\left({1-f\over g_0}\right)^{1\over 2}\right] ^{f}
\phi^{-{3\,f\over 2}}.
$$
 In this context, the condition of
inflation to occur is given by  $\varepsilon_1<$1, or analogously
$\ddot{a}>0$. Therefore, the inflaton field during the
inflationary epoch is such that $\phi>\left(
{\frac{1-f}{Af}}\right) ^{{2\over 3\,f}} \left[ \frac{3}{2}
\left({1-f\over g_0}\right)^{1\over 2}\right] ^{{2\over 3}}$.
As we mentioned earlier,  the inflationary phase begins at the
earliest possible stage, i.e,
$\varepsilon_1(\phi=\phi_1)=1$. Then, the scalar field $\phi_{1}$,
is given by $ \phi_{1}=\left( {\frac{1-f}{Af}}\right) ^{{2\over
3\,f}} \left[ \frac{3}{2} \left({1-f\over g_0}\right)^{1\over
2}\right] ^{{2\over 3}}$. Also the number of $e$-folds $N$ defined
between two different values of cosmological times $t_{1}$ and
$t_{2}$ or equality between $\phi_{1}$ and $\phi_{2}$, by considering
Eq.(\ref{wr11})  can be written as \begin{equation}
N=\int_{t_{1}}^{t_{2}}\,H\,dt=A\,\left(
t_{2}^{f}-t_{1}^{f}\right)
=\frac{2^{f}A}{3^{f}} \left({g_0\over 1-f}\right)^{f\over 2}\,\left(  \phi_{2}^{3f/2}-\phi_{1}^{3f/2}\right)  .
\label{N1}%
\end{equation}

From the number of e-folding $N$, it is possible to rewrite the
function ${\mathcal{A}}=3g_0H\dot{\phi}$ in terms of $N$. Thus,  from
Eqs.(\ref{at}), (\ref{gfi1}) and (\ref{N1}), we have that
\begin{equation}
\mathcal{A}(N)=2^{1\over 3}(3g_0)^{2\over 3}(Af)(1-f)^{1\over 3}
\left[{Af\over 1+f(N-1) } \right]^{4-3f\over 3f}. \label{awr}%
\end{equation}

Since the cosmological perturbations depend on the dissipation
coefficient $\Gamma$, then  in the   following we will analyze our
model in the limit ${\mathcal{A}}\gg1+R$, for two specific cases
of the dissipation coefficient $\Gamma$ studied in the literature, namely;
$\Gamma(\phi)=\Gamma_0=$ constant \cite{warm1,warm2} and
$\Gamma(\phi)\propto\,V(\phi)$\cite{delCampo:2007cy}.

\subsection{Case $\Gamma=\Gamma_0=$ constant.}

Let us consider that our model of G-warm inflation evolves
according to the
regime ${\mathcal{A}}\gg1+R$, when
the
dissipation coefficient $\Gamma$ has the following form, where $\Gamma=\Gamma_0=$
constant\cite{warm1,warm2}. In this sense, from Eq.(\ref{PRy2})
we find that the power spectrum of the scalar perturbations
${\mathcal{P}_{\mathcal{R}}}$, can be rewritten as
 \begin{equation}
{\mathcal{P}}_{\mathcal{R}}={\Gamma_0^{1/4}  P_0\over 3^{1/ 4} }
\,
 H^{43\over 12}
 (-\dot{H})^{-{1\over 3}},\,\,\,\,\mbox{where}\,\,\,\,P_0=\frac{3^{19/12}\,g_0^{5/6}}{2^{4/3}\pi^2\,C_\gamma^{1/4}}.
\end{equation}
Here, we have used Eq.(\ref{gfi1}). Now, by using Eq.(\ref{wr11}), we
can write the  power spectrum of the scalar perturbation in terms
of the inflaton field as
\begin{equation}
{\mathcal{P}_{\mathcal{R}}}(\phi)=P_{I}\,\phi^{-\beta_{I}},\,\;\;\;\mbox{in
which}\;\;\;\;P_{I}=P_{0} \left({\Gamma_0\over 3} \right) ^{1\over
4} \left(Af\right)^{39\over 12}\left(1-f\right)^{-1\over 3}\left[
\frac{3}{2} \left({1-f\over g_0}\right)^{1\over 2}\right]
^{2\beta_I\over 3},
\end{equation}
and  $\beta_I$ is defined as $\beta_{I}=\left[{35-39f\over
8}\right]$. Note that for the particular case in which
$f=35/39\simeq0.90$, the power spectrum of the scalar perturbations
becomes constant. From Eq.(\ref{N1}),
 we can rewrite  the   power
spectrum of the scalar perturbation as a function of the number of
$e$-folds $N$ as
\begin{equation}
{\mathcal{P}_{\mathcal{R}}}(N)=p_{I}\,\left[{Af\over 1+f(N-1) }
\right]^{2\beta_I ,\over 3f},\label{PP1NN}
\end{equation}
where the constant $p_I$ is defined as $p_{I}=P_{0}
\left({\Gamma_0\over 3} \right) ^{1\over 4}
\left(Af\right)^{13\over 4}\left(1-f\right)^{-1\over 3}$.

As the scalar spectral index $n_{s}$ is defined  as
$n_{s}-1=\frac{d\ln \,{\mathcal{P}_{R}}}{d\ln k}$, we find that the index
$n_s$ can be written in terms of the scalar field $\phi$ as
\begin{equation}
n_{s}=1-\left( {35-39f\over 12Af}\right) \left[ \frac{3}{2}
\left({1-f\over g_0}\right)^{1\over 2}\right] ^{f}
\phi^{-{3\,f\over 2}} .
\label{nss2}%
\end{equation}
Also, we note that for the specific value of $f= 35/39\simeq
0.90$, the scalar spectral index $n_s$ corresponds to a
scale-invariant spectral index, for which $n_s=1$, called the
Harrison-Zel'dovich spectrum of density perturbations. As  we
mentioned before,  for intermediate inflation  in the context of
GR, the
 parameter $f=2/3$  corresponds to the value $n_s=1$.
From Eq.(\ref{N1}),  we also obtain the
scalar spectral index $n_s$  as function of $N$, yielding
\begin{equation}
n_{s}=1-{35-39f\over 12[1+f(N-1)]}.
\label{nswr}%
\end{equation}
Note that from this equation we can express the parameter $f$ in terms of the spectral index and the
number of $e$-folds as $f=\frac{12(n_s-1)+35}{3[13+4(1-n_s)(N-1)]}$. In
particular, for the number of e-folds $N=60$ and the scalar spectral index $n_s=0.967$, we find that the
value of the parameter $f$ is given by $f\simeq 0.55$. Also,  for $N=60$ and considering the current observational constraint for $n_s$ set by
Planck, given by $n_s=0.964$, the parameter $f$ corresponds to $f\simeq0.54$.

 Furthermore, we can express the
parameter $A$ of the intermediate expansion in terms of the quantities  $g_0$, $\Gamma_0$, $N$,
${\mathcal{P}_{\mathcal{R}}}(N)$ and $f$ (or equivalently $n_s$)  as
\begin{equation}
A=\left[\frac{3^{1/4}{\mathcal{P}_{\mathcal{R}}}}{f^{13/4}P_0\,\Gamma_0^{1/4}}
(1-f)^{1/3}
\left(\frac{1+f(N-1)}{f}\right)^{2\beta_I/3f}\right]^{\frac{12f}{39f+8\beta_I}}.\label{Aa1}
\end{equation}
Here, we have considered Eq.(\ref{PP1NN}).

From Eq.(\ref{te}), the tensor-to-scalar ratio $r$ as a function  of the
scalar spectral index $n_s$ can be written as

\begin{equation}
r(n_s) \simeq {2A^2f^2\over \pi^2\,p_I} \left[ {35-39f\over
12Af(1-n_s)}\right] ^{11-15f\over 12f}.\label{nsok}
\end{equation}

We also mention that the ratio $R=\Gamma/3H$ can be expressed   as
a function of  the number of $e$-folds by considering  Eq.(\ref{N1}). In doing so, we have that the ratio $R=R(N)$  becomes
\begin{equation}
R(N)={\Gamma_{0}\over 3Af}\left[{1+f(N-1) \over Af}
\right]^{1-f\over f}.\label{N8}
\end{equation}
 Similarly, from   Eqs.(\ref{awr}), (\ref{nswr}) and (\ref{N8}), we can obtain
the effective function ${\mathcal{A}}-R$ in terms of the scalar spectral
index $n_s$, resulting
\begin{equation}
{\mathcal{A}}-R=[2(1-f)]^{1/3}(3g_0)^{2/3}(Af)^{4/3f}\left[\frac{12(1-n_s)}{35-39f}\right]^{\frac{4-3f}{3f}}-\frac{\Gamma_0}{3(Af)^{1/f}}
\left[\frac{35-39f}{12(1-n_s)}\right]^{\frac{1-f}{f}}.\label{AA1}
\end{equation}

Note that in order to achieve
the domination of the Galileon coupling during the whole inflationary
stage,  we must take into account that ${\mathcal{A}}\gg 1+R$.
\begin{figure}[th]
{{\hspace{-2.3cm}\includegraphics[width=2.9in,angle=0,clip=true]{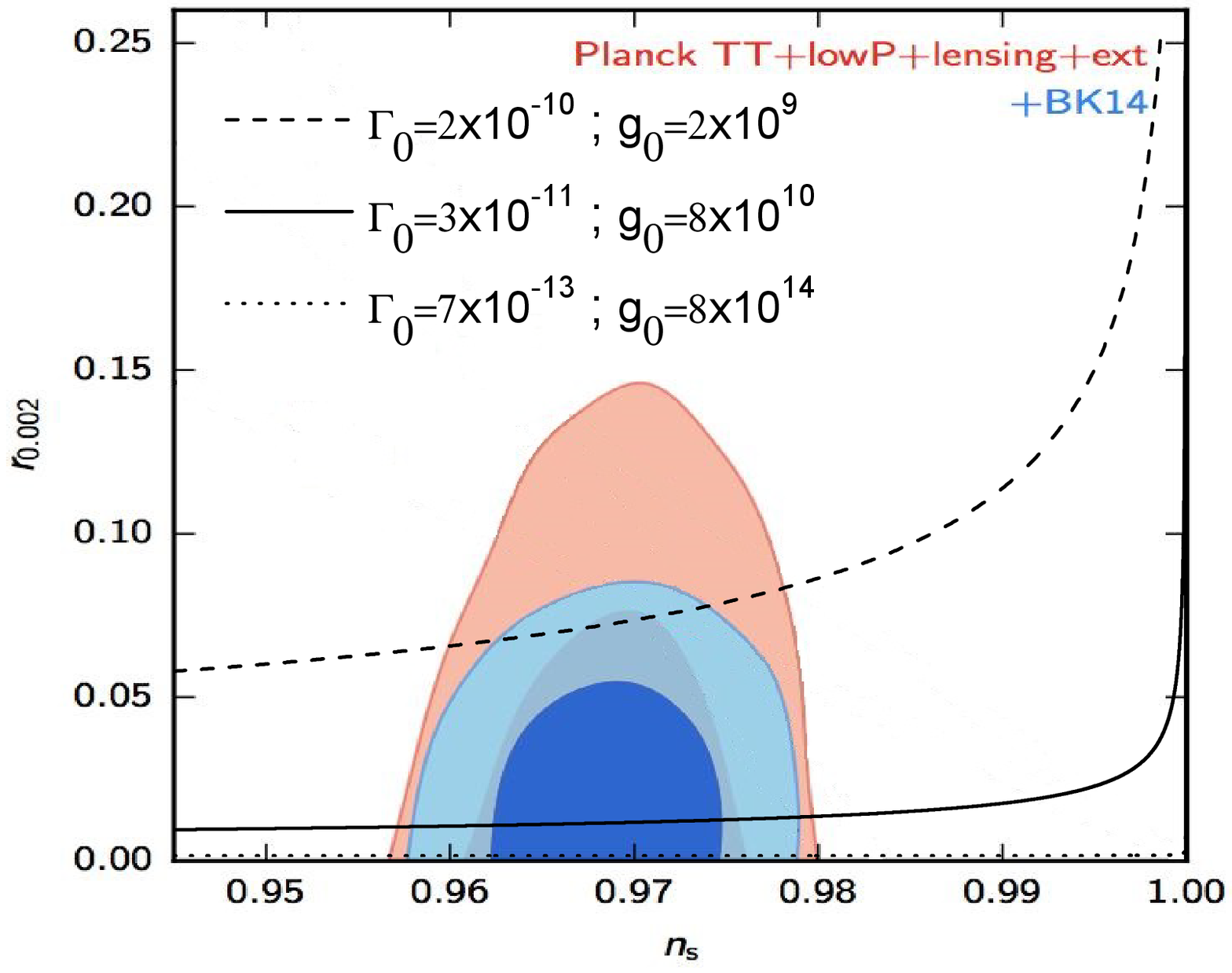}}}
{{\hspace{1.9cm}\includegraphics[width=3.9in,angle=0,clip=true]{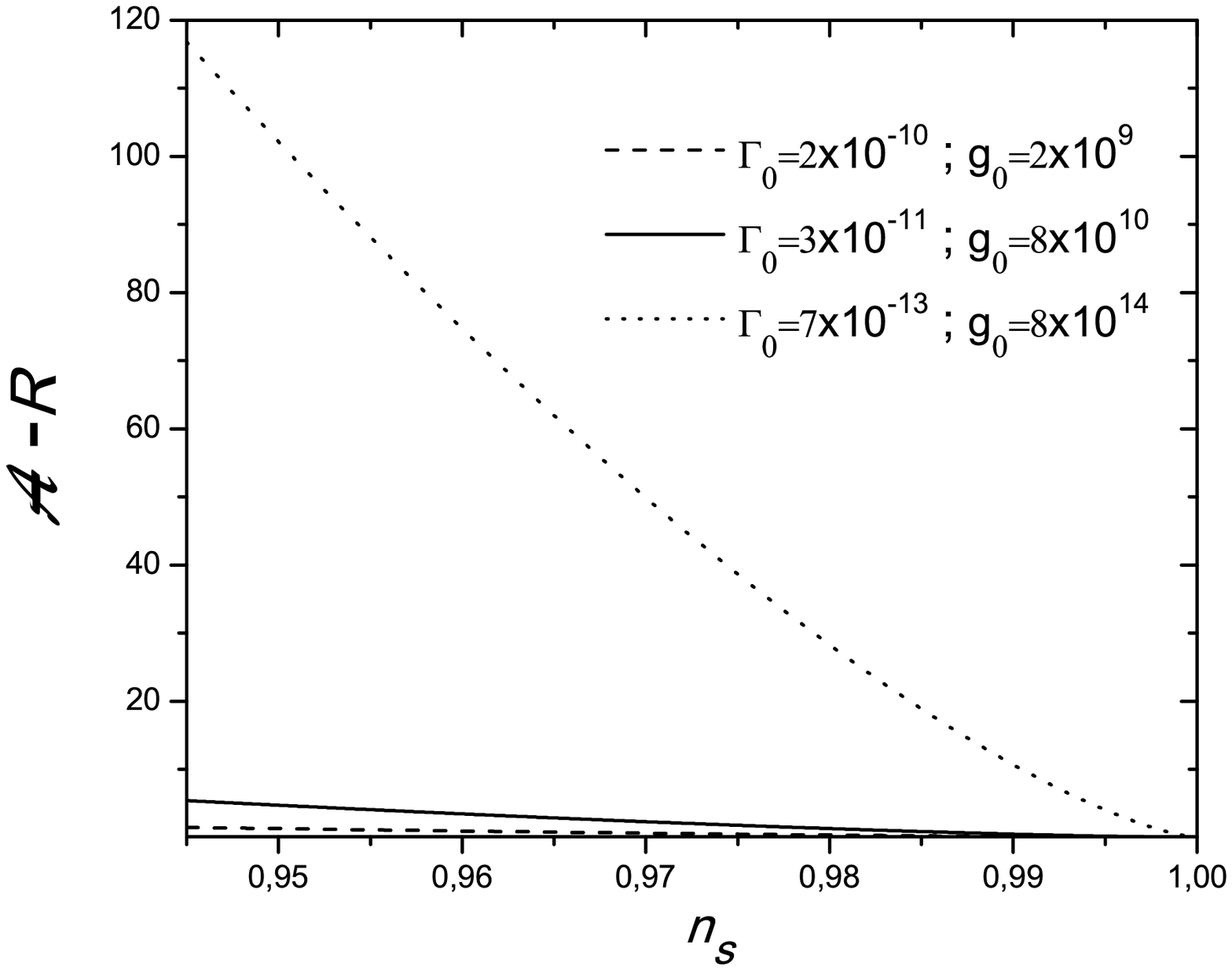}}}
{\vspace{-0.8cm}\caption{ Plot of the tensor-to-scalar ratio $r$
against the scalar spectral index $n_s$(upper panel), and the the
difference ${\mathcal{A}}-R$ as a functions of scalar spectral
index $n_s$(lower panel), in the G-warm intermediate model when
$\Gamma=\Gamma_0=$ const. For both panels we  use three a
different pair of values of ($\Gamma_0,g_0$).
 \label{fig01}}}
\end{figure}

Also, the temperature of the thermal bath can be rewritten   from
Eq.(\ref{temp}) as
\begin{equation}
T=\left[ \frac{\Gamma_0}{4C_{\gamma}}\right] ^{1/4} \left[
\frac{2}{3g_0}\right] ^{1/6} H^{-5/12} (-\dot{H})^{1/6},
 \label{t1b}%
\end{equation}
and from Eqs.(\ref{H1}),(\ref{nss2}) and (\ref{t1b}) the rate
$T/H$ in terms of the scalar spectral index $n_s$ can be written
as
\begin{equation} \frac{T}{H}(n_s)=\left[ \frac{\Gamma_0}{4C_{\gamma}}\right]
^{1/4} \left[ \frac{2}{3g_0}\right] ^{1/6} {(1-f)^{1/6}\over
(Af)^{13/12f}} \left[ {35-39f\over 12(1-n_s)}\right] ^{13-15f\over
12f}>1.
\label{t10b1}%
\end{equation}
Here, we have considered that the essential condition for warm  inflation to occur, is set by $T>H$ \cite{warm1,warm2}.

Fig.\ref{fig01} shows the tensor-to-scalar ratio $r$ versus
the scalar spectral index $n_s$ (upper panel) and in the lower
panel we show the  necessary condition for domination of the
Galileon term in which ${\mathcal{A}}-R\gg1$ versus the scalar
spectral index $n_s$, when $\Gamma=\Gamma_0=$ constant. For both plots, we have
considered three different pairs of values  $(\Gamma_0,g_0)$.  In the
upper panel are shown the two-dimensional marginalized constraints at
68$\%$ and 95$\%$ confide level on the consistency relation
$r=r(n_s)$ from Ref.\cite{Array:2015xqh}. The
 lower panel shows the dependence of the difference between  the function
 ${\mathcal{A}}$ and the rate $R$ on the scalar spectral index, and we ensure that the condition of domination Galileon effect in our model be valid ,i.e. ${\mathcal{A}}\gg
 1+R$. For  the upper plot
   we
use Eq.(\ref{nsok}) in order to obtain the consistency relation $r=r(n_s)$. Also, in order to write down values that associate the difference of
 ${\mathcal{A}}-R$ with the scalar spectral index $n_s$, we
 considered Eq.(\ref{AA1}) (lower panel). On the other hand, to get the
 pair $(g_0,\Gamma_0)$, we have manipulated numerically Eqs.(\ref{nsok}) and  (\ref{t10b1}), the
 form to the satisfy the  essential condition for warm  inflation $T/H>1$, and the observational constraint on the
 consistency relation, given by $r=r(n_s)<0.07$. From these relations, the lower bounds for the parameters are are found to be $g_0>2\times 10^{9}$ and $\Gamma_0>2\times
 10^{-10}$. Here, we have used Eqs.(\ref{nswr}) and (\ref{Aa1}) together with the number
 of $e$-folds set to $N=60$. Analogously,  for the specific case in which $T/H>1$ and
 $r=r(n_s)<0.01$ we obtained that $g_0>8\times 10^{10}$ and $\Gamma_0>3\times
 10^{-12}$. Also,  for the case $T/H>1$ and
 $r=r(n_s)<0.0001$ we found  that the pair of  parameters $(g_0,\Gamma_0)$ have as lower limits;  $g_0>8\times 10^{14}$ and $\Gamma_0>7\times
 10^{-13}$, respectively. However, from the lower plot we find
 that for lower  bounds  $g_0>8\times 10^{14}$ and $\Gamma_0>7\times
 10^{-13}$, the G-warm model evolves according to the regime of domination of the Galilean, for which ${\mathcal{A}}\gg1+R$, for the intermediate expansion when
 $\Gamma(\phi)=\Gamma_0=$constant. However, for the  limits of $g_0>8\times 10^{14}$ and $\Gamma_0>7\times
 10^{-13}$, we noted that  the tensor-to-scalar ratio is such that  $r\sim
 0$. In this sense, the observational data from the consistency relation $r=r(n_s)$ does
 not impose constraints on the parameter-space.
 Lastly, for the case in which the coefficient $\Gamma=\Gamma_0=$ constant, we find that the constraint
 for the parameter $f$ associated to intermediate scale factor is
 given by $f\simeq 0.55$ and the constraints for the parameter
 $g_0$ and $\Gamma_0$ are found to be $g_0>8\times 10^{14}$ and $\Gamma_0>7\times
 10^{-13}$, respectively.


\subsection{Case $\Gamma(\phi)\propto V(\phi)$.}
Following Ref.\cite{delCampo:2007cy}, we consider that the
dissipative coefficient in terms of the scalar field
$\Gamma(\phi)$ is given by $\Gamma(\phi)=k\,V(\phi)$, where $k>0$
corresponds to a constant.
 By considering  Eq.(\ref{PRy2}), we obtain that the power
spectrum of the scalar perturbation ${\mathcal{P}_{\mathcal{R}}}$,
in the limit  ${\mathcal{A}}\gg1+R$ becomes

\begin{equation}
 {\mathcal{P}_{\mathcal{R}}}=k ^{1\over 4} P_0\, H^{49\over
12}
(-\dot{H})^{-{1\over 3}}. \label{pdab}%
 \end{equation}
As before, we can find the   power spectrum of the scalar
perturbation in terms  of the number of e-folds $N$ as
\begin{equation}
{\mathcal{P}_{\mathcal{R}}}(N)=p_{II}\,\left[{Af\over 1+f(N-1) } \right]^{41-45f\over 12f}, \label{pd3}%
\end{equation}
with $p_{II}$ defined as $p_{II}=P_{0}\,
k^{1\over 4} \left(Af\right)^{15\over 4}\left(1-f\right)^{-1\over
3}$.  Also, we find that  the scalar spectral index $n_s=n_s(\phi)$ becomes
\begin{equation}
n_{s}=1-\left( {41-45f\over 12Af}\right) \left[ \frac{3}{2}
\left({1-f\over g_0}\right)^{1\over 2}\right] ^{f} \phi^{-{3\,f\over
2}} ,
\label{nss2f}%
\end{equation}
or, in terms of the number of $e$-folds this results in
\begin{equation}
n_{s}=1-{41-45f\over 12[1+f(N-1)]}. \label{nswr2}%
\end{equation}
Here, we have used Eq.(\ref{N1}). Again, we observe that for the
special value of $f= 41/45\simeq 0.91$, we have $n_s=1$, yielding the
Harrison-Zel'dovich spectrum of density perturbations. As before,
we realize that we may express the parameter $f$ in terms of the
scalar spectral index as well as the number of $e$-folds as
$f=[12(n_s-1)+45]/[12(N-1)(1-n_s)+45]$. In particular, setting $N=60$
and considering the maximum likelihood value for $n_s$ found by Planck 2015 \cite{Ade:2015lrj},
given by $n_s=0.967$,
we obtain that $f$ has the value $f\simeq 0.59$. Now for the current observational value$n_s=0.964$ \cite{Akrami:2018odb}, we found that $f\simeq 0.58$.
From Eq.(\ref{pd3}), we can express the parameter
$A$ as a function of the parameters $g_0$, $k$, $N$ and $f$ as follows
\begin{equation}
A=\left(\frac{{\mathcal{P}_{\mathcal{R}}}(1-f)^{1/3}}{P_0
k^{1/4}\,f^{15/4}}\right)^{12f/41}\,\left[\frac{1+f(N-1)}{f}\right]^{\frac{41-45f}{41}}.\label{ATT}
\end{equation}

By considering Eq.(\ref{te}), the tensor-to-scalar ratio $r$, written in terms of  the
scalar spectral index $n_s$ becomes
\begin{equation}
r (n_s)\simeq {2A^2f^2\over \pi^2\,p_{II}} \left[ {41-45f\over
12Af(1-n_s)}\right] ^{17-21f\over 12f} .\label{rns2b2}
\end{equation}
Analogously to the case of $\Gamma=\Gamma_0=$ constant, we note
that the ratio $R=\Gamma/3H$ can be expressed in terms of the
number of $e$-folding $N$, from Eq.(\ref{N1}), as
\begin{equation}
R(N)=k\left[{Af\over 1+f(N-1) } \right]^{1-f\over f}.\label{Ratio-ns2}%
\end{equation}

Also as before, we can express the difference ${\mathcal{A}}-R$ as
function of the scalar spectral index $n_s$, yielding
\begin{equation}
{\mathcal{A}}-R=[2(1-f)]^{1/3}(3g_0)^{2/3}(Af)^{4/3f}\left[\frac{12(1-n_s)}{41-45f}\right]^{\frac{4-3f}{3f}}-k(Af)^{1/f}
\left[\frac{12(1-n_s)}{41-45f}\right]^{\frac{1-f}{f}}.\label{arns2}
\end{equation}
Here we have used Eqs.(\ref{awr}), (\ref{nswr2})
and(\ref{Ratio-ns2}).

On the other hand, from Eq.(\ref{temp}), the temperature of the thermal bath can be rewritten as follows
\begin{equation} T=\left[ \frac{3k}{4C_{\gamma}}\right]
^{1/4} \left[ \frac{2}{3g_0}\right] ^{1/6} H^{1/12}
(-\dot{H})^{1/6},
\label{t2b}%
\end{equation}
and from Eqs.(\ref{H1}),(\ref{nss2f}) and (\ref{t2b}) the ratio
$T/H$ as in terms of the scalar spectral index $n_s$, becomes
\begin{equation} \frac{T}{H}(n_s)=\left[ \frac{3k}{4C_{\gamma}}\right]
^{1/4} \left[ \frac{2}{3g_0}\right] ^{1/6} {(1-f)^{1/6}\over
(Af)^{7/12f}} \left[ {41-45f\over 12(1-n_s)}\right] ^{7-9f\over
12f}>1.
\label{t10b2}%
\end{equation}
Recall that the essential condition for warm  inflation to occur
is such that $T/H>1$.

\begin{figure}[th]
{{\hspace{-2.3cm}\includegraphics[width=2.9in,angle=0,clip=true]{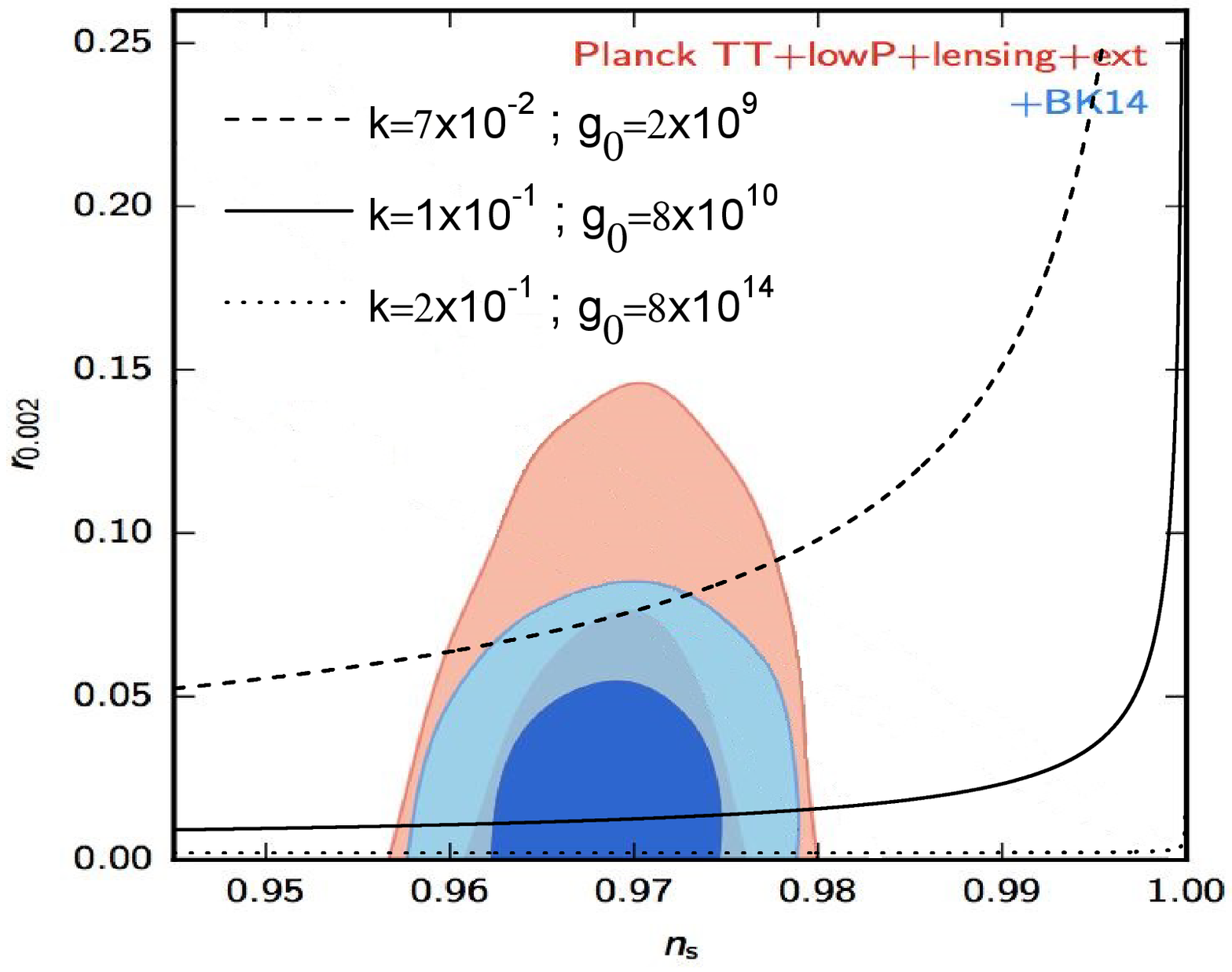}}}
{{\hspace{1.9cm}\includegraphics[width=3.9in,angle=0,clip=true]{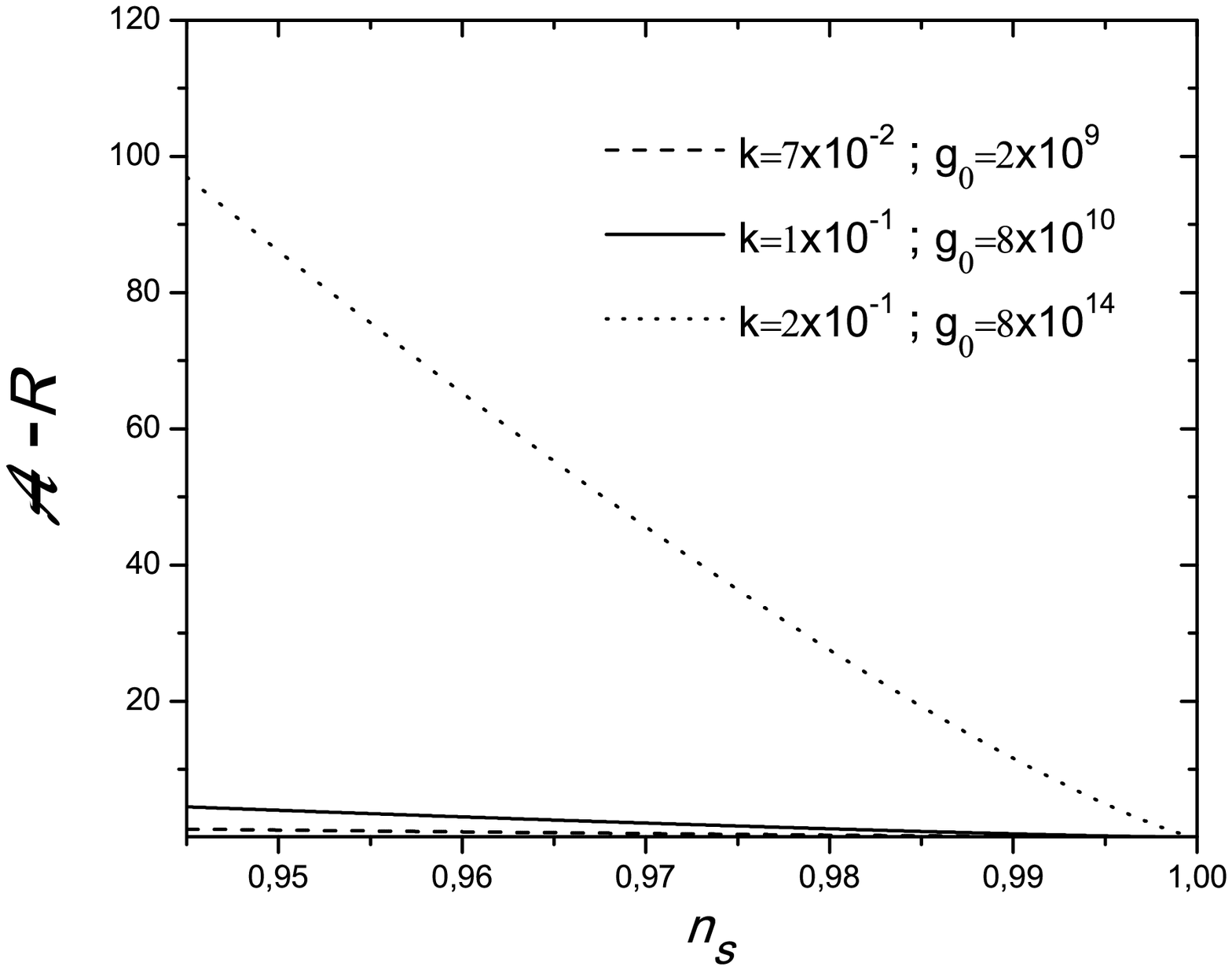}}}
{\vspace{-0.5 cm}\caption{  The evolution of the tensor to scalar
ratio $r$ versus the scalar spectral index $n_s$(upper panel) and
the evolution of the difference ${\mathcal{A}}-R$ versus the
scalar spectral index $n_s$(lower panel) in the G-warm
intermediate model for the case in which the dissipative
coefficient depends of the scalar field as $\Gamma(\phi)\propto
V(\phi)$. In both panels we use three different values of the
pairs ($k,g_0$).
 \label{fig02}}}
\end{figure}

In the upper panel of Fig.\ref{fig02}, we plot the tensor-to-scalar ratio
$r$ against the scalar spectral index $n_s$, and in
the lower panel we show the  necessary condition of domination of
the Galileon effect in which ${\mathcal{A}}\gg1+R$ versus the
scalar spectral index $n_s$, in the case in which the dissipation
coefficient $\Gamma(\phi)\propto V(\phi)$. For both panels, we have
considered three different pairs $(k,g_0)$.  The upper panel shows
the two-dimensional marginalized constraints at 68$\%$ and 95$\%$
C.L. on the consistency relation $r=r(n_s)$. The
 lower panel shows the evolution of the difference
 ${\mathcal{A}}-R$ during the inflationary scenario. Here, we make sure that the
 condition of domination Galileon effect  in which  ${\mathcal{A}}\gg
 1+R$ is valid.
 In the upper panel  we
consider the consistency relation $r=r(n_s)$ from
Eq.(\ref{rns2b2}). Also, in order to write down values that associate
the difference
 ${\mathcal{A}}-R$ to the scalar spectral index $n_s$, we
 considered Eq.(\ref{arns2}) (lower panel). To obtain the
 pair $(k,g_0)$, we numerically solve Eqs.(\ref{rns2b2}) and (\ref{t10b2}),   in order to
 satisfy  the  constraint on the
 consistency relation $r=r(n_s)<0.07$ as well as the  essential condition for warm  inflation to
 occur, $T/H>1$. In this way, the constraints on
 the several parameters are found to be $g_0>2\times 10^{9}$ and $k>7\times
 10^{-2}$. Here, we have used Eqs.(\ref{ATT}) for the value of $A$  together with the number of $e$-folds $N=60$.
 Analogously  as before, for the specific case in which $T/H>1$ and
 $r=r(n_s)<0.01$, we obtained that the lower limit for   $g_0>8\times 10^{10}$ and $k_0>1\times
 10^{-1}$. Similarly, for the special case in which $T/H>1$ and
 $r=r(n_s)<0.0001$, we found  that the lower bounds for the pair of the parameters are given by  $g_0>8\times 10^{14}$ and $k>2\times
 10^{-1}$, respectively. Here, it is worth to mention that the lower bound for the
 parameter $g_0$ is similar to the case in which the dissipative
 coefficient is $\Gamma_0=$const.

 As  before, from the lower plot we observe
 that for  $g_0>8\times 10^{14}$ and $k>2\times
 10^{-1}$, the G-warm model evolves according to the domination
 of the Galilean coupling, i.e. ${\mathcal{A}}\gg1+R$. Similarly as before, we noted that for the pair $g_0>8\times 10^{14}$ and $k>2\times
 10^{-1}$, the G-warm model is able to predict a tensor-to-scalar ratio such that $r\sim 0$. In fact,
  in order to satisfy the condition of
 domination of Galileon coupling, given by ${\mathcal{A}}\gg1+R$, we have that $r\sim 0$. In this sense,   the consistency relation $r=r(n_s)$ does
 not impose any constraints on the space of parameters as the previous case.

\section{ General solution. \label{section41}}

In this section we will study the general solution of G-warm
intermediate inflationary model. In this sense, we will consider
that the left terms of Eq.(\ref{campo}) are similar i.e.,
$R\sim{\mathcal{A}}\sim 1$, that  we will  call it the general solution.
From the slow-roll equation of motion for the inflaton field  given by
Eq.(\ref{campo}), we can obtain an equation for $\dot{\phi}$ given
by \be \dot{\phi}^3+\left({1+R\over 3\,g_0H}\right)\dot{\phi}^2
-{2(-\dot{H})\over 3\,g_0H}=0. \,\label{gfi} \en Here we note that
this equation depends on the ratio $R=\Gamma/3H$. Thus,  in   the
following we will analyze our model  for two specific cases of the
dissipation coefficient $\Gamma$. The first case we will analyze
corresponds to $\Gamma(\phi)=\Gamma_0=$ constant
 and in the second case we will study the case in which
$\Gamma(\phi)\propto V(\phi)$, as it was previously studied.

\subsection{Case $\Gamma=\Gamma_0=$ constant.}
Let us consider that our model of G-warm inflation takes place for
constant dissipative coefficient $\Gamma=\Gamma_0$ during the
regime in which ${\mathcal{A}}\sim R\sim 1$. From
Eq.(\ref{gfi})  we find that the speed of the scalar field $\dot{\phi}$ can be written as
\begin{equation}
\dot{\phi}={3H+\Gamma_0\over 27g_0H^2}\left[-1+ 2
\cosh\left({1\over 3} \cosh^{-1}\left[{ 3^8g_0^2H^5(-\dot{H})\over
(3H+\Gamma_0)^3}-1\right] \right) \right].
 \label{wr1e}%
\end{equation}

From Eq.(\ref{PRy2}) the power spectrum of the scalar perturbation
results
\begin{equation}
{\mathcal{P}_{\mathcal{R}}}={\sqrt{3}\over 2\pi^2 } \left(
{\Gamma_0\over 4C_{\gamma}} \right) ^{1\over 4}
 H ^{11\over 4}
\dot{\phi}^{-{3\over 2}} \sqrt{1+\Gamma_0/3H+6g_0H\dot{\phi}},
 \label{pd}%
\end{equation}
and  since the scalar spectral index $n_{s}$ is given by
$n_{s}-1=\frac{d\ln
    \,{\mathcal{P}_{R}}}{d\ln k}$,  we have

\begin{equation}
n_{s}=1-{11\over 4}\varepsilon_1+{3\over 2}\epsilon_2^{(I)}+
{1\over 2}\epsilon_5^{(I)},\label{nsGo}
\end{equation}
where the coefficient $\epsilon_2^{(I)}$ is given by
$$
\epsilon_2^{(I)}={2\ddot{H}+\dot{\phi}^2\dot{R}+3g_0\dot{H}\dot{\phi}^3\over
2H\dot{\phi}^2(1+R)+9g_0H^2\dot{\phi}^3},
$$
and the parameter $\epsilon_5^{I}$ is defined as
$$
\epsilon_5^{(I)}={-\Gamma_0\dot{H}/3H^2
+6g_0\dot{H}\dot{\phi}+6g_0H\ddot{\phi}\over
H(1+\Gamma_0/3H+6g_0H\dot{\phi})},\,\,\,\mbox{with}\,\,\,\,\,-\ddot{\phi}={2\ddot{H}+\dot{\phi}^2\dot{R}+3g_0\dot{H}\dot{\phi}^3\over
2\dot{\phi}(1+R)+9g_0H\dot{\phi}^2}.
$$
Here $\dot{R}=-\Gamma_0H^{-2}\dot{H}/3$.

Recall that the Hubble rate in terms of the number of $e$-folds $N$
for intermediate inflation  can be rewritten as $
H(N)=Af\,\left[{Af\over 1+f(N-1) } \right]^{1-f\over f}, $ and
also $ -\dot{H}(N)=Af(1-f)\,\left[{Af\over 1+f(N-1) }
\right]^{2-f\over f}, $  see Eqs.(\ref{HN}) and (\ref{DHN}),
respectively. Then, we may express both the power spectrum of the scalar
perturbation ${\mathcal{P}_{\mathcal{R}}}$ and the scalar spectral
index $n_s$ can in terms of $N$, or similarly as a function of the Hubble rate $H(N)$ in the
form $
{\mathcal{P}_{\mathcal{R}}}={\mathcal{P}_{\mathcal{R}}}[H(N)]$,
and  $n_s=n_s[H(N)],
 $
respectively.

Also from Eq.(\ref{te}), we may write the tensor-to-scalar ratio $r$,
for the full solution when $\Gamma=\Gamma_0=$ constant. In this form, we have

\begin{equation}
r \simeq {4\over \sqrt{3} } \left( {4C_{\gamma}\over \Gamma_0 }
\right) ^{1\over 4} H ^{-3\over 4} \dot{\phi}^{{3\over 2}}
(1+\Gamma_0/3H+6g_0H\dot{\phi})^{-1/2}
,\label{Rk11}%
\end{equation}
where $\dot{\phi}$ is given by Eq.(\ref{wr1e}). As before, the
tensor-to-scalar ratio $r$ can be rewritten in terms of the number
of $e$-folds $N$ as $r=r[H(N)]$.
\begin{figure}[th]
{{\hspace{-2.3cm}\includegraphics[width=2.9in,angle=0,clip=true]{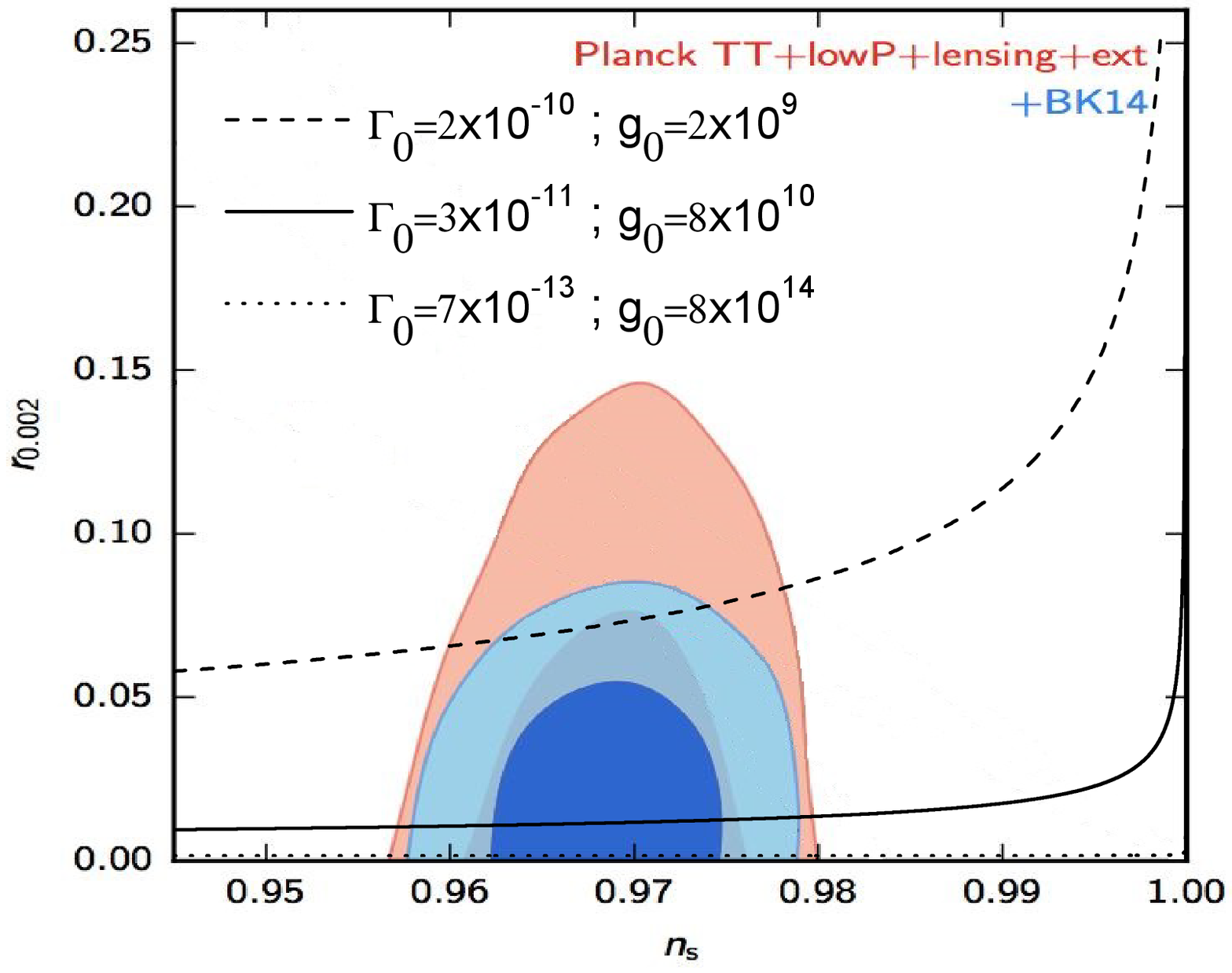}}}
{{\hspace{1.9cm}\includegraphics[width=3.9in,angle=0,clip=true]{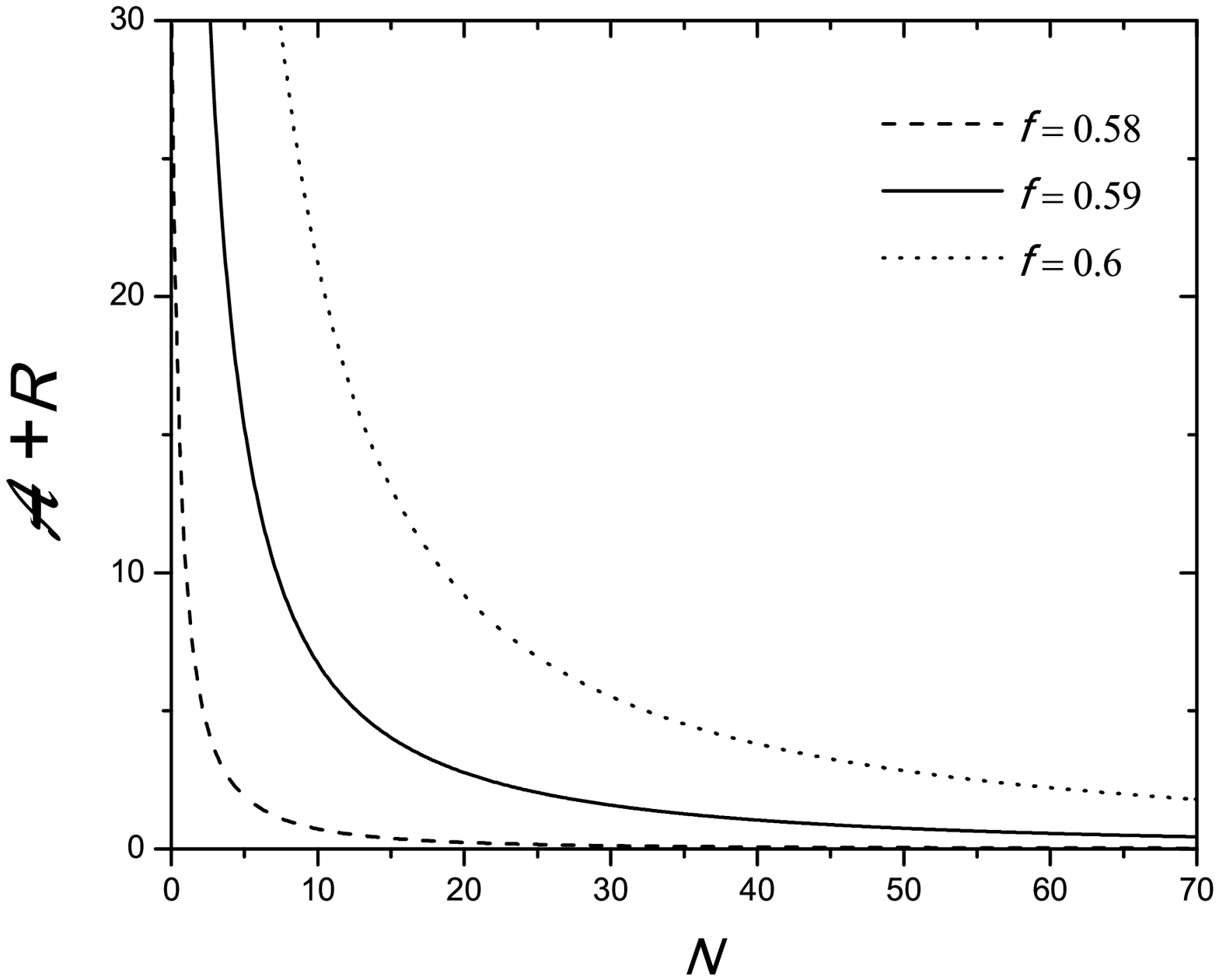}}}
{\vspace{-0.5 cm}\caption{  Plot of the tensor-to-scalar ratio $r$
against the scalar spectral index $n_s$(upper
panel)\cite{Array:2015xqh} and the evolution of the function
${\mathcal{A}}+R$ versus the number of $e$-folds $N$(lower panel)
in the G-warm intermediate model for $\Gamma=\Gamma_0=$constant,
for the general solution. In both panels we use three several
values of the parameter $f$ with their corresponding trios of
values $(\Gamma_0,g_0,A)$.
 \label{fig03}}}
\end{figure}

In Fig.\ref{fig03} we show the plot of the tensor-to-scalar ratio $r$ against the scalar spectral index $n_s$ (upper panel). Here, we show the
two-dimensional marginalized constraints at 68$\%$ and 95$\%$ C.L.
on the consistency relation $r=r(n_s)$ from  BICEP2/Keck Array
Collaborations data\cite{Array:2015xqh}. In the lower panel, we
show ${\mathcal{A}}+R$ as a function of the number of
$e$-folds $N$. In particular, it is
  depicted the evolution of the function
 ${\mathcal{A}}+R$ during the inflationary period i.e.,  between the number of $e$-folds $N=0$ (beginning of inflation, see Eq.(\ref{Nf})) and $N=70$.
We also establish  that the
 condition in which ${\mathcal{A}}\sim R\sim 1$, is satisfied, in order to be consistent with  the full solution to the Klein-Gordon equation, see Eq.(\ref{campo})
 (under slow roll approximation).
In both panels we considered  the case when $\Gamma=\Gamma_0=$ constant, and we have also fixed
three different values of $f$, which characterizes the intermediate
expansion law.

In order to write down values that relate $r$ and $n_s$, we
numerically  manipulate Eqs.(\ref{nsGo}) and (\ref{Rk11}) to get the
consistency relation $r=r(n_s)$  (upper plot).
 Analogously, to relate
 the effective function
 ${\mathcal{A}}+R$ to the number of $e$-folds $N$ between $N=0$ to $N=70$ during the inflationary stage, we
 numerically
 utilize Eqs.(\ref{HN}), (\ref{DHN}) and (\ref{wr1e}), see  the lower
 panel. In order to obtain the trio of  parameters $(\Gamma_0,g_0,A)$ for fixed value of parameter $f$, which characterizes the intermediate expansion law ,
 we consider the last data Planck collaboration \cite{Akrami:2018odb}, which set the power spectrum of the scalar
perturbation to ${\mathcal{P}_{\mathcal{R}}}\simeq 2.2\times
10^{-9}$, and the scalar spectral index to $n_s\simeq0.964$, and we also consider the
minimum   condition for warm  inflation to occur, $T/H=1$ . Here, we have  fixed the number of $e$-folds to $N=60$. In this sense, the
corresponding trio of values $(\Gamma_0,g_0,A)$ for $f= 0.58$, is found to be $(4.9\times
10^{-11}, 2.7\times 10^{8}, 1.2\times 10^{-2})$. Analogously for
the value $f=0.59$, we obtained $(2.3\times
10^{-11}, 1.4\times 10^{10}, 8.6\times 10^{-3})$. In a similarly fashion, for
$f=0.6$ we
 determined that the trio of values is given by $(7.9\times
10^{-12}, 2.6\times 10^{11}, 5.4\times 10^{-3})$.

 From the lower  panel  of
Fig.\ref{fig03}, we observe that in order to satisfy the condition
${\mathcal{A}}\sim R\sim 1$ given by the full  Klein-Gordon
equation (see (\ref{campo})), we obtain that the upper limit for
the parameter $f$ is given by $f<0.6$. In this context, we note
that for values of $f>0.6$, the effective function
${\mathcal{A}}+R\gg 0$, during the inflationary epoch, and the
model does not evolves in agreement to the general regime ${\mathcal{A}}\sim R\sim 1$. However, from  the upper panel
 we note that
 the upper  bound for $f$ is given by $f<0.6$, since   the model is well supported by
 the Planck data from the consistency relation $r=r(n_s)$. Here,
 both conditions are satisfied. We also mentioned that, according to
 the parameter $f$ increases, the corresponding values for the
parameters $\Gamma_0$ and $A$ decrease, however the parameter
$g_0$ increase.

\subsection{Case $\Gamma(\phi)\propto V(\phi)$}

Now we assume  that our G-model of warm inflation takes place for
dissipative coefficient being a function of the scalar field $\phi$
given by  $\Gamma(\phi)=kV(\phi)$, during the regime in with
${\mathcal{A}}\sim R\sim 1$, i.e. the full Klein-Gordon equation
(\ref{campo}) under slow-roll approximation. In this way, from Eq.(\ref{gfi}) we find that
$\dot{\phi}$ can be written as
\begin{equation}
\dot{\phi}={1+kH\over 9g_0H}\left[-1+ 2 \cosh\left({1\over 3}
\cosh^{-1}\left[{ 3^5g_0^2H^2(-\dot{H})\over (1+kH)^3}-1\right]
\right) \right].
\label{wr1rb}%
\end{equation}

For this dissipative coefficient, the power spectrum of the scalar
perturbation ${\mathcal{P}_{\mathcal{R}}}$, yields
\begin{equation}
{\mathcal{P}}_{\mathcal{R}}={\sqrt{3}\over 2\pi^2 } \left(
{3k\over 4C_{\gamma}} \right) ^{1\over 4} H ^{13\over 4}
\dot{\phi}^{-{3\over 2}} \sqrt{1+kH+6g_0H\dot{\phi}}.
\label{pd4}%
\end{equation}

Thus, we obtain that the scalar spectral index $n_s$  results in
\begin{equation}
n_{s}=1-{13\over 4}\varepsilon_1+{3\over 2}\epsilon_2^{(II)}+
{1\over 2}\epsilon_5^{(II)},
\label{nss2d}%
\end{equation}
where $\epsilon_2^{(II)}$ is defined as
$$
\epsilon_2^{(II)}=\epsilon_2^{(I)}={2\ddot{H}+\dot{\phi}^2\dot{R}+3g_0\dot{H}\dot{\phi}^3\over
2H\dot{\phi}^2(1+R)+9g_0H^2\dot{\phi}^3},\,\,\,\mbox{with}\,\,\,\,\dot{R}=k\dot{H},
$$
and the parameter $\epsilon_5^{(II)}$ is given by
$$
\epsilon_5^{(II)}={k\dot{H}+6g\dot{H}\dot{\phi}+6gH\ddot{\phi}\over
H(1+kH+6g_0H\dot{\phi})}.
$$
Here $\dot{\phi}$ corresponds to Eq.(\ref{wr1rb}) and
$\ddot{\phi}$ is given by
$\ddot{\phi}=-\left[{2\ddot{H}+\dot{\phi}^2\dot{R}+3g_0\dot{H}\dot{\phi}^3\over
2\dot{\phi}(1+R)+9g_0H\dot{\phi}^2}\right]$.

As before,  we find that the tensor-to-scalar ratio $r$, for the full
solution when
$\Gamma(\phi)\propto V(\phi)$ becomes
\begin{equation}
r \simeq {4\over \sqrt{3} } \left( {4C_{\gamma}\over 3k} \right)
^{1\over 4} H ^{-5\over 4} \dot{\phi}^{{3\over 2}}
(1+kH+6g_0H\dot{\phi})^{-1/2}.
\label{Rk11b}%
 \end{equation}
Here we have used  Eq.(\ref{te}). As in the previous case, we can
rewrite the  power spectrum of the scalar perturbation
${\mathcal{P}_{\mathcal{R}}}$,  the scalar spectral index $n_s$
and the tensor-to-scalar ratio $r$  in terms of the number of $e$-folds $N$, or
similarly as a function of the Hubble rate $H(N)$ in the form
$ {\mathcal{P}_{\mathcal{R}}}={\mathcal{P}_{\mathcal{R}}}[H(N)]$,
  $n_s=n_s[H(N)]$ and $r=r[H(N)]$.

\begin{figure}[th]
{{\hspace{-2.3cm}\includegraphics[width=2.9in,angle=0,clip=true]{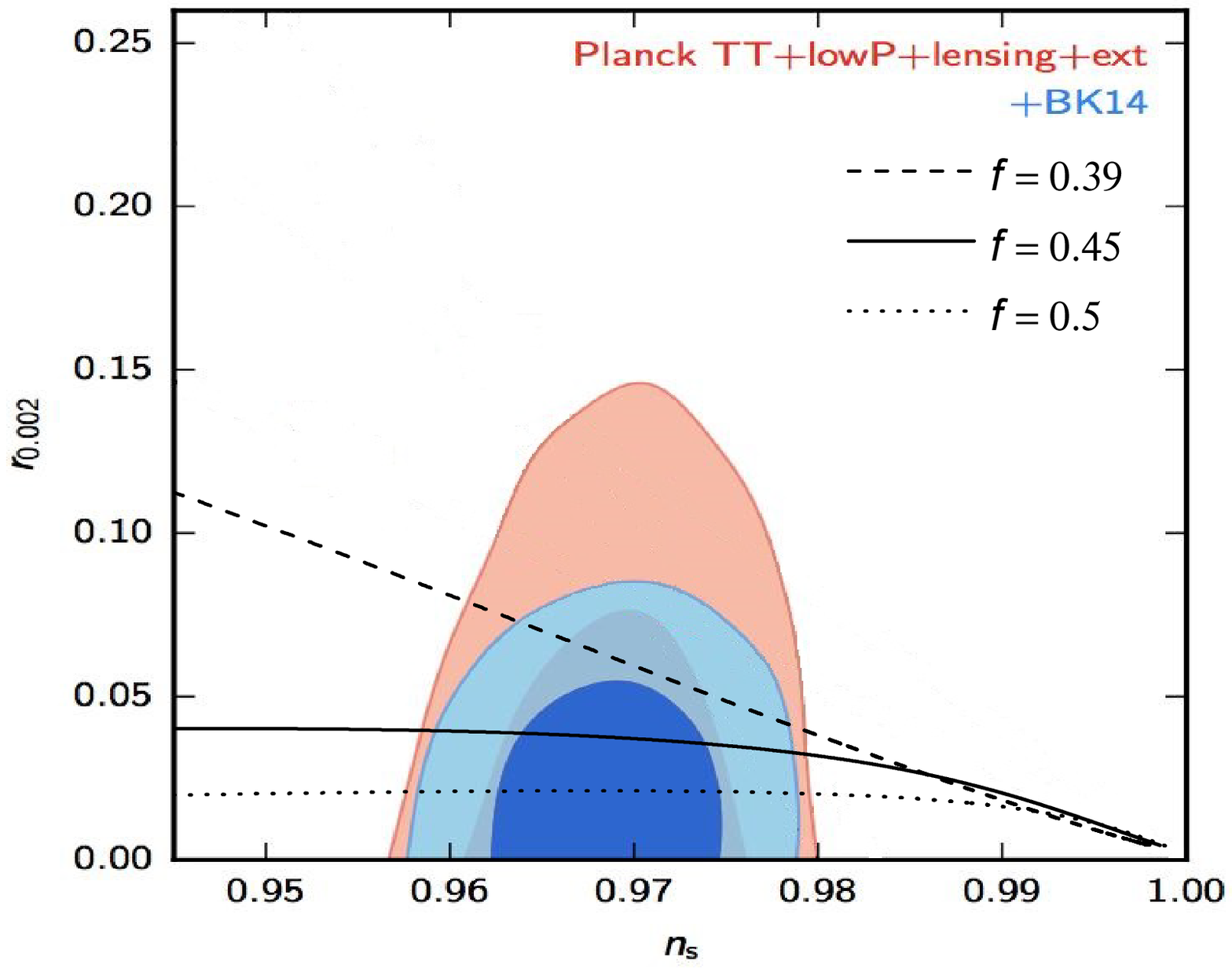}}}
{{\hspace{1.9cm}\includegraphics[width=3.9in,angle=0,clip=true]{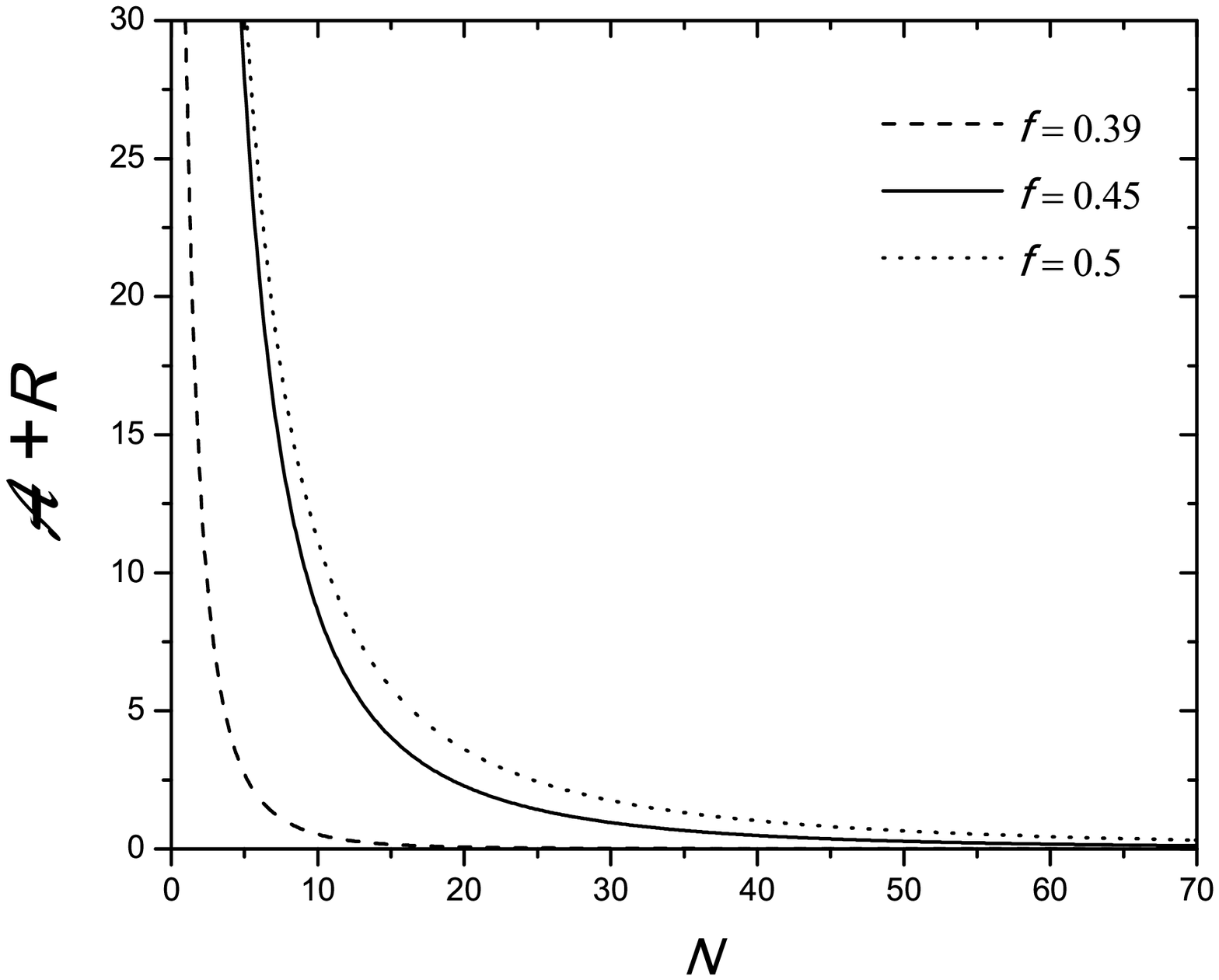}}}
{\vspace{-0.5 cm}\caption{  The tensor-to-scalar ratio $r$ as a
function of the scalar spectral index
$n_s$\cite{Array:2015xqh}(upper panel) and the evolution of the
${\mathcal{A}}+R$ in terms the number of $e$-folds $N$(lower
panel) in the G-warm intermediate model  when $\Gamma(\phi)\propto
V(\phi)$ for the general solution. In both panels we use three
different values of the parameter $f$.
 \label{fig04}}}
\end{figure}

Analogously as before, in Fig.\ref{fig04} we show the tensor-to-scalar ratio $r$ versus the scalar spectral index $n_s$ (upper
panel). Here, we show the two-dimensional marginalized constraints
at 68$\%$ and 95$\%$ C.L. on the consistency relation $r=r(n_s)$
from Ref.\cite{Array:2015xqh}. In the lower panel we show the
function ${\mathcal{A}}+R$ versus the number of $e$-folds $N$. In
this
  panel we exhibit  the evolution of the function
 ${\mathcal{A}}+R$ during the inflationary period between the number of $e$-folds $N=0$ and $N=70$.
We also check that the
 condition ${\mathcal{A}}\sim R\sim 1$ is satisfied, in order to obtain the full expression  to the Klein-Gordon equation (\ref{campo})
 under slow-roll approximation.
In both panels we considered  that  $\Gamma(\phi)\propto V(\phi)$ as well as
three different values of the parameter $f$ .

As before, by manipulating numerically Eqs.(\ref{nss2d}) and
(\ref{Rk11b}), we obtain the consistency relation $r=r(n_s)$ for
the upper plot.
 Analogously, for
 the function
 ${\mathcal{A}}+R$ versus the number of $e$-folds $N$, we
 numerically
 considered Eqs.(\ref{HN}), (\ref{DHN}) and (\ref{wr1rb}) in order to plot ${\mathcal{A}}+R$ against $n_s$ (lower panel).

 Since the parameter $f$ lies in the range $0<f<1$, we fixed the
 valuer of $f$,
 in order to obtain the
trio of values $(k,g_0,A)$. Then, we numerically utilize
Eqs.(\ref{temp}),(\ref{pd4}) and (\ref{nss2d})  to satisfy the minimum   condition for that warm  inflation takes place
  in which $T/H=1$, the power spectrum of the scalar
perturbation ${\mathcal{P}_{\mathcal{R}}}=2.2\times 10^{-9}$ and
the scalar spectral index $n_s=0.964$ for a given value of $f$. In
particular, by fixing the number of $e$-folds to $N=60$, together with
$T/H(N=60)=1$, ${\mathcal{P}_{\mathcal{R}}}(N=60)=2.2\times
10^{-9}$, $n_s(N=60)=0.964$ and $f$=0.39, we find numerically that the trio of
values of $(k,g_0,A)$ is given by $(0.5, 3.3\times 10^{6}, 0.3)$.
Analogously, for $f=0.45$, we obtained numerically
 the trio $(0.6,8.7\times 10^{8},0.1)$. Similarly, for $f=0.5$ we
 determined that the trio corresponds to $(0.9,5.2\times 10^{9},4.1\times10^{-2})$.

  From the upper panel  of Fig.\ref{fig04}, we observe that
 the upper bound for $f$ becomes $f<0.39$, since   the model is well supported by
 the Planck data in $n_s-r$ plane. However, from the lower
 panel we note that in order to satisfy the condition  ${\mathcal{A}}\sim R\sim 1$ (in the full  Klein-Gordon equation
(\ref{campo})), the upper limit for the parameter $f$
is found to be $f<0.5$. In this context, we determine that for values of
$f>0.5$, the effective function becomes ${\mathcal{A}}+R\gg 0$ during inflation, hence the model does not evolves according to
the condition ${\mathcal{A}}\sim R\sim 1$. Numerically, we also noted if
the parameter $f$ increases, both the associated  parameters with the
dissipative coefficient, $k$ and the coupling parameter $g_0$
increase, while the associated parameter to the intermediate
expansion $A$ decreases. It is interesting to highlight that the
allowed ranges for the parameters $f$, $k$, $g_0$ and $A$ for the full
model are found only from the condition in which the full-model evolves according to ${\mathcal{A}}\sim R\sim
1$. In this form, we find that the consistency relation $r=r(n_s)$ does not impose any constraints on the parameters for  this model.

\section{Conclusions \label{conclu}}

In this paper we have investigated the realization of the
intermediate inflationary
model in G-warm inflation scenario. By assuming the Galileon
term under the slow roll-approximation, we have considered the
coupling function as $G(\phi,X)=g_0 X$, where $g_0=$
constant, for two different dissipation coefficients in the
scenario of intermediate warm inflation. In particular, we have
studied two expressions for the dissipative coefficient, namely $\Gamma=\Gamma_0=$ constant
and $\Gamma(\phi)\propto V(\phi)$. In addition, we have
assumed that the dynamics takes place according two regimes. In the first one, we have considered the
domination of the Galilean coupling over the standard terms of warm
inflation. In the second regime, we have considered that
all terms become of the same
order in the slow-roll equation for the scalar field. By assuming the intermediate expansion law, we have found
analytical solutions to the background equations under the
slow-roll approximation for each regime, considering
the two expressions for the dissipative coefficient. Also, for both regimes, we
have found the constraints on the several parameters, assuming the
last data of Planck in addition to the condition of domination term
associated with its regime.

In order to developed the analysis for the first regime, or   domination of the
Galileon term i.e. ${\cal{A}}\gg 1+\Gamma/3H$, we have set the
parameter $f$ from the expression for scalar spectral index and the parameter $A$
from the amplitude of the power spectrum of scalar perturbations. In order to obtain the parameters
characterizing the coupling $G(\phi,X)$ and the dissipative coefficient
$\Gamma$, such as  $g_0$ and $\Gamma_0$ (or the pair $(g_0,k)$),
 we have solved numerically the conditions for warm inflation, i.e. $T>H$ and the consistency relation
$r=r(n_s)<0.07$ from last data of Planck. Thus, for the regime in which the
domination of warm inflation comes from the  Galilean coupling, we
have obtained the constraints on the parameters of our model, which only come
from the condition ${\cal{A}}\gg 1+\Gamma/3H$,
giving a lower bound on the parameter-space.

In this sense,  the
consistency relation $r=r (n_s )$ does not impose any constraints on
the parameters, since the tensor-to-scalar ratio $r\sim 0$ for the
allowed range of parameters.
 We have  found that the
lower bound on the
 parameter $g_0$ is similar to the  different types of   dissipation
 coefficients;  $\Gamma=\Gamma_0=$constant and $\Gamma\propto V(\phi) $ during for regime in
 which ${\cal{A}}\gg 1+\Gamma/3H$.

In the second stage of the analysis of our model, we consider the dynamics
takes place in the so-called general regime of Eq.(\ref{gfi}) (considering slow-roll
approximation). Here, we have fixed the parameter $f$ associated to the
intermediate expansion $f$ which lies in the range $0<f<1$. Also, in order to find the other
parameters, such $A$, from the intermediate expansion law, the coupling of $G(\phi,X)$
and the ones which characterize the dissipative coefficient $\Gamma$, namely $g_0$ and
$\Gamma_0$ (or the pair $(g_0,k)$), we have solved numerically the
conditions for warm inflation in which the temperature $T=H$, and
the consistency relation in which $r=r(n_s)<0.07$ from last data of Planck. For the several expression for the dissipative coefficient,  we have found that
the current observational data of Planck does not impose any constraints on
the space of parameters. On the other hand, we have found that only the
condition for the model evolves according to $R\sim 1\sim
\cal{A}$ is able to impose the constraints on the parameters characterizing our model.
In this sense, we have found that
 these models are well
supported by the last Planck data , since the tensor-to-scalar ratio $r<0.07$. Also, due to the difficulty in treating the
equations analytically, we have the study of this regime (general solution)
in numerical way.

As a final remark, we have not studied G-warm inflation  in the framework of intermediate expansion when the coupling function $g$ has a dependence on the inflaton, as neither
a  dissipative coefficient having a dependence on the temperature of
the thermal bath $T$, i.e., $\Gamma(\phi,T)$. We hope
to be able to address these points in a future work.

\begin{acknowledgments}
R.H. was supported by Proyecto VRIEA-PUCV N$_{0}$ 039.309/2018.
N.V. acknowledges support from the Fondecyt de Iniciaci\'on
project N$^o$ 11170162.
\end{acknowledgments}



\begin{thebibliography}{99}                                                                                               %




























\bibitem {R1}A. Guth , Phys. Rev. D \textbf{23}, 347 (1981).

\bibitem {R102}A.A. Starobinsky, Phys. Lett. B \textbf{91}, 99
(1980).

\bibitem {R106}K. Sato, Mon. Not. Roy. Astron. Soc. \textbf{195}, 467 (1981).

\bibitem {R103}A.D. Linde, Phys. Lett. B \textbf{108}, 389 (1982).

\bibitem {R104}A.D. Linde, Phys. Lett. B \textbf{129}, 177 (1983).

\bibitem {R105}A. Albrecht and P. J. Steinhardt, Phys. Rev. Lett.
\textbf{48},1220 (1982).


\bibitem {R2}V.F. Mukhanov and G.V. Chibisov , JETP Letters \textbf{33},
532(1981).

\bibitem {R202}S. W. Hawking,Phys. Lett. B \textbf{115}, 295 (1982).

\bibitem {R203}A. Guth and S.-Y. Pi, Phys. Rev. Lett. \textbf{49}, 1110 (1982).

\bibitem {R204}A. A. Starobinsky, Phys. Lett. B \textbf{117}, 175 (1982).

\bibitem {R205}J.M. Bardeen, P.J. Steinhardt and M.S. Turner, Phys. Rev.D
\textbf{28}, 679 (1983).


\bibitem{Aghanim:2018eyx}
  N.~Aghanim {\it et al.} [Planck Collaboration],
  arXiv:1807.06209 [astro-ph.CO].

\bibitem{Ade:2015xua}
  P.~A.~R.~Ade {\it et al.} [Planck Collaboration],
  Astron.\ Astrophys.\  {\bf 594}, A13 (2016).

\bibitem{Ade:2013zuv}
  P.~A.~R.~Ade {\it et al.} [Planck Collaboration],
  Astron.\ Astrophys.\  {\bf 571}, A16 (2014).


\bibitem{Akrami:2018odb}
  Y.~Akrami {\it et al.} [Planck Collaboration],
  arXiv:1807.06211 [astro-ph.CO].

\bibitem{Array:2015xqh}
  P.~A.~R.~Ade {\it et al.} [BICEP2 and Keck Array Collaborations],
  Phys.\ Rev.\ Lett.\  {\bf 116}, 031302 (2016).

  \bibitem{Ade:2015lrj}
  P.~A.~R.~Ade {\it et al.} [Planck Collaboration],
  Astron.\ Astrophys.\  {\bf 594}, A20 (2016).

  \bibitem{Planck:2013jfk}
  P.~A.~R.~Ade {\it et al.} [Planck Collaboration],
  Astron.\ Astrophys.\  {\bf 571}, A22 (2014).


\bibitem{Kofman:1997yn}
  L.~Kofman, A.~D.~Linde and A.~A.~Starobinsky,
  Phys.\ Rev.\ D {\bf 56}, 3258 (1997).

\bibitem{Kofman:1994rk}
  L.~Kofman, A.~D.~Linde and A.~A.~Starobinsky,
  Phys.\ Rev.\ Lett.\  {\bf 73}, 3195 (1994).


  \bibitem{Amin:2014eta}
  M.~A.~Amin, M.~P.~Hertzberg, D.~I.~Kaiser and J.~Karouby,
  Int.\ J.\ Mod.\ Phys.\ D {\bf 24}, 1530003 (2014).

\bibitem{Allahverdi:2010xz}
  R.~Allahverdi, R.~Brandenberger, F.~Y.~Cyr-Racine and A.~Mazumdar,
  Ann.\ Rev.\ Nucl.\ Part.\ Sci.\  {\bf 60}, 27 (2010).



\bibitem{warm1}
I.G. Moss, Phys.Lett.B \textbf{154}, 120 (1985).
A. Berera,   Phys. Rev. Lett. {\bf 75}, 3218 (1995).

\bibitem{warm2}
A. Berera, Phys. Rev. D {\bf 55}, 3346 (1997).



\bibitem{BasteroGil:2012cm}
  M.~Bastero-Gil, A.~Berera, R.~O.~Ramos and J.~G.~Rosa,
  JCAP {\bf 1301}, 016 (2013).

  \bibitem{Bartrum:2013fia}
  S.~Bartrum, M.~Bastero-Gil, A.~Berera, R.~Cerezo, R.~O.~Ramos and J.~G.~Rosa,
  Phys.\ Lett.\ B {\bf 732}, 116 (2014).

\bibitem{Zhang:2009ge}
  Y.~Zhang,
  JCAP {\bf 0903}, 023 (2009).

\bibitem {26}I.~G.~Moss and C.~Xiong,
arXiv:hep-ph/0603266.

\bibitem {28}A.~Berera, M.~Gleiser and R.~O.~Ramos,
Phys.\ Rev.\ D \textbf{58} 123508 (1998).

\bibitem {PRD} J. Yokoyama and A. Linde, Phys. Rev D {\bf 60},
083509, (1999).



\bibitem {6252602}I.G. Moss, Phys.Lett.B \textbf{154}, 120 (1985).

\bibitem {1126}A. Berera, Phys. Rev.D \textbf{54}, 2519 (1996).

\bibitem {6252603}A. Berera and L.Z. Fang, Phys.Rev.Lett. \textbf{74} 1912
(1995).

\bibitem {6252604}A. Berera, Nucl.Phys B \textbf{585}, 666 (2000).

\bibitem {62526} L.M.H. Hall, I.G. Moss and A. Berera, Phys.Rev.D \textbf{69},
083525 (2004).

\bibitem{Moss:2008yb} I.~G.~Moss and C.~Xiong,
  JCAP {\bf 0811}, 023 (2008).

\bibitem{Ramos:2013nsa} R.~O.~Ramos and L.~A.~da Silva,
  JCAP {\bf 1303}, 032 (2013).

\bibitem {Berera:2008ar} A.~Berera, I.~G.~Moss and R.~O.~Ramos,
Rept.\ Prog.\ Phys.\ \textbf{72}, 026901 (2009); \\
M.~Bastero-Gil
and A.~Berera,
Int.\ J.\ Mod.\ Phys.\ A \textbf{24}, 2207 (2009).

\bibitem{Ramos:2016coz} R.~O.~Ramos,
  Astrophys.\ Space Sci.\ Proc.\  {\bf 45}, 283 (2016).


\bibitem{Das:2018rpg}
  S.~Das,
  arXiv:1810.05038 [hep-th].

\bibitem{Motaharfar:2018zyb}
  M.~Motaharfar, V.~Kamali and R.~O.~Ramos,
  arXiv:1810.02816 [astro-ph.CO].

 \bibitem{Bastero-Gil:2018uep}
  M.~Bastero-Gil, A.~Berera, R.~Hernández-Jiménez and J.~G.~Rosa,
  Phys.\ Rev.\ D {\bf 98}, no. 8, 083502 (2018).

 \bibitem{Li:2018wno}
  X.~B.~Li, H.~Wang and J.~Y.~Zhu,
  Phys.\ Rev.\ D {\bf 97}, no. 6, 063516 (2018).


  \bibitem{Herrera:2018cgi}
  R.~Herrera,
  Eur.\ Phys.\ J.\ C {\bf 78}, no. 3, 245 (2018).





\bibitem{Lucchin:1984yf}
  F.~Lucchin and S.~Matarrese,
  Phys.\ Rev.\ D {\bf 32}, 1316 (1985).

\bibitem{Barrow:1990vx}
  J.~D.~Barrow,
  Phys.\ Lett.\ B {\bf 235}, 40 (1990).

\bibitem{Barrow:1993zq}
  J.~D.~Barrow and A.~R.~Liddle,
  Phys.\ Rev.\ D {\bf 47}, no. 12, R5219 (1993).

  \bibitem{Barrow:2006dh}
  J.~D.~Barrow, A.~R.~Liddle and C.~Pahud,
  Phys.\ Rev.\ D {\bf 74}, 127305 (2006).



  \bibitem{Kamali:2016frd} S.~del Campo and R.~Herrera,
  JCAP {\bf 0904}, 005 (2009);
  V.~Kamali, S.~Basilakos and A.~Mehrabi,
  Eur.\ Phys.\ J.\ C {\bf 76}, no. 10, 525 (2016).

  \bibitem{Herrera:2016sov} S.~del Campo and R.~Herrera,
  Phys.\ Lett.\ B {\bf 653}, 122 (2007); S.~del Campo and R.~Herrera,
  Phys.\ Lett.\ B {\bf 670}, 266 (2009);
  R.~Herrera, N.~Videla and M.~Olivares,
  Eur.\ Phys.\ J.\ C {\bf 76}, no. 1, 35 (2016).

  \bibitem{Herrera:2015aja} S.~del Campo, R.~Herrera and A.~Toloza,
  Phys.\ Rev.\ D {\bf 79}, 083507 (2009);
  R.~Herrera, N.~Videla and M.~Olivares,
  Eur.\ Phys.\ J.\ C {\bf 75}, no. 5, 205 (2015).

  \bibitem{Herrera:2014nta} R.~Herrera and N.~Videla,
  Eur.\ Phys.\ J.\ C {\bf 67}, 499 (2010);
  R.~Herrera, N.~Videla and M.~Olivares,
  Phys.\ Rev.\ D {\bf 90}, no. 10, 103502 (2014).

  \bibitem{Herrera:2014mca} R.~Herrera and E.~San Martin,
  Eur.\ Phys.\ J.\ C {\bf 71}, 1701 (2011); R.~Herrera and E.~San Martin,
  Int.\ J.\ Mod.\ Phys.\ D {\bf 22}, 1350008 (2013);
  R.~Herrera, M.~Olivares and N.~Videla,
  Int.\ J.\ Mod.\ Phys.\ D {\bf 23}, no. 10, 1450080 (2014).

  \bibitem{Herrera:2013rra}
  R.~Herrera, M.~Olivares and N.~Videla,
  Phys.\ Rev.\ D {\bf 88}, 063535 (2013); C.~Gonzalez and R.~Herrera,
  Eur.\ Phys.\ J.\ C {\bf 77}, no. 9, 648 (2017).



  \bibitem{Horndeski:1974wa}
  G.~W.~Horndeski,
  Int.\ J.\ Theor.\ Phys.\  {\bf 10}, 363 (1974).

  \bibitem{Nicolis:2008in}
  A.~Nicolis, R.~Rattazzi and E.~Trincherini,
  Phys.\ Rev.\ D {\bf 79}, 064036 (2009).

  \bibitem{Deffayet:2011gz}
  C.~Deffayet, X.~Gao, D.~A.~Steer and G.~Zahariade,
  Phys.\ Rev.\ D {\bf 84}, 064039 (2011).

  \bibitem{Kobayashi:2011nu}
  T.~Kobayashi, M.~Yamaguchi and J.~Yokoyama,
  Prog.\ Theor.\ Phys.\  {\bf 126}, 511 (2011).

  \bibitem{Charmousis:2011bf}
  C.~Charmousis, E.~J.~Copeland, A.~Padilla and P.~M.~Saffin,
  Phys.\ Rev.\ Lett.\  {\bf 108}, 051101 (2012).








\bibitem{DeFelice:2011hq}
  A.~De Felice, T.~Kobayashi and S.~Tsujikawa,
  Phys.\ Lett.\ B {\bf 706}, 123 (2011).

  \bibitem{Tsujikawa:2014rta}
  S.~Tsujikawa,
  PTEP {\bf 2014}, no. 6, 06B104 (2014).

  \bibitem{Ohashi:2012wf}
  J.~Ohashi and S.~Tsujikawa,
  JCAP {\bf 1210}, 035 (2012).


  \bibitem{Herrera:2018mvo}
  R.~Herrera,
  Phys.\ Rev.\ D {\bf 98}, no. 2, 023542 (2018).

  \bibitem{Ramirez:2018dxe}
  H.~Ramirez, S.~Passaglia, H.~Motohashi, W.~Hu and O.~Mena,
  JCAP {\bf 1804}, no. 04, 039 (2018).

  \bibitem{Maity:2018ipt}
  D.~Maity and P.~Saha,
  JCAP {\bf 1807}, no. 07, 065 (2018).

  \bibitem{Hirano:2016gmv}
  S.~Hirano, T.~Kobayashi and S.~Yokoyama,
  Phys.\ Rev.\ D {\bf 94}, no. 10, 103515 (2016).

  \bibitem{Unnikrishnan:2013rka}
  S.~Unnikrishnan and S.~Shankaranarayanan,
  JCAP {\bf 1407}, 003 (2014).




  \bibitem{Kamada:2010qe}
  K.~Kamada, T.~Kobayashi, M.~Yamaguchi and J.~Yokoyama,
  Phys.\ Rev.\ D {\bf 83}, 083515 (2011).

  \bibitem{Kobayashi:2010cm}
  T.~Kobayashi, M.~Yamaguchi and J.~Yokoyama,
  Phys.\ Rev.\ Lett.\  {\bf 105}, 231302 (2010).

  \bibitem{Burrage:2010cu}
  C.~Burrage, C.~de Rham, D.~Seery and A.~J.~Tolley,
  JCAP {\bf 1101}, 014 (2011).


 \bibitem{BazrafshanMoghaddam:2016tdk}
  H.~Bazrafshan Moghaddam, R.~Brandenberger and J.~Yokoyama,
  Phys.\ Rev.\ D {\bf 95}, no. 6, 063529 (2017).

  \bibitem{Herrera:2017qux}
  R.~Herrera,
  JCAP {\bf 1705}, no. 05, 029 (2017).

  \bibitem{Motaharfar:2017dxh}
  M.~Motaharfar, E.~Massaeli and H.~R.~Sepangi,
  Phys.\ Rev.\ D {\bf 96}, no. 10, 103541 (2017).


  \bibitem{Teimoori:2017jzo}
  Z.~Teimoori and K.~Karami,
  Astrophys.\ J.\  {\bf 864}, no. 1, 41 (2018).

  \bibitem{Herrera:2018ker}
  R.~Herrera, N.~Videla and M.~Olivares,
  arXiv:1806.04232 [gr-qc].




  \bibitem{nw1} X.~M.~Zhang, H.~Y.~Ma, P.~C.~Chu, J.~T.~Liu and J.~Y.~Zhu,
  JCAP {\bf 1603}, no. 03, 059 (2016); P.~Goodarzi and H.~Mohseni Sadjadi,
  arXiv:1609.06185 [gr-qc].

  \bibitem{nw2}M.~Sharif and A.~Ikram,
  J.\ Exp.\ Theor.\ Phys.\  {\bf 123}, no. 1, 40 (2016); M.~Jamil, D.~Momeni and R.~Myrzakulov,
  Int.\ J.\ Theor.\ Phys.\  {\bf 54}, no. 4, 1098 (2015);
 X.~M.~Zhang and j.~Y.~Zhu,
  Phys.\ Rev.\ D {\bf 90}, no. 12, 123519 (2014).

\bibitem{sh} S.~del Campo and R.~Herrera,
  Phys.\ Rev.\ D {\bf 76}, 103503 (2007); S.~del Campo, R.~Herrera, J.~Saavedra, C.~Campuzano and E.~Rojas,
  Phys.\ Rev.\ D {\bf 80}, 123531 (2009).

  \bibitem{delCampo:2007cy}
  S.~del Campo, R.~Herrera and D.~Pavon,
  Phys.\ Rev.\ D {\bf 75}, 083518 (2007);  M.~R.~Setare and V.~Kamali,
  arXiv:1312.2832 [physics.gen-ph]; A.~Cid,
  Phys.\ Lett.\ B {\bf 743}, 127 (2015).



\end{thebibliography}
\end{document}